%% file: GAMA-VVF.tex
\documentclass[useAMS,usenatbib]{mnras} 
\usepackage{graphicx}
\usepackage{amssymb, amsmath}
\usepackage{times}
\usepackage[dvipsnames]{xcolor}
\usepackage{hyperref}
\hypersetup{colorlinks=true,allcolors=teal}
\usepackage{todonotes}
\usepackage{lscape}
\usepackage{booktabs}
\usepackage[section]{placeins}
\usepackage{breqn}
\usepackage{setspace}
\usepackage{adjustbox}

\usepackage{lineno}

\usepackage{arydshln}
\usepackage{url}
\usepackage{bbold}

\definecolor{carrotorange}{rgb}{0.93, 0.57, 0.13}

\newcommand{\gtsima}{$\; \buildrel > \over \sim \;$}
\newcommand{\ltsima}{$\; \buildrel < \over \sim \;$}
\newcommand{\simgt}{\lower.7ex\hbox{\gtsima}}
\newcommand{\simlt}{\lower.7ex\hbox{\ltsima}}

\newcommand{\Mpch}{\ensuremath{h^{-1}{\rm Mpc}}}

\newcommand{\Mh}{\ensuremath{h^{-1}M_{\odot}}}
\newcommand{\avg}[1]{\ensuremath{\left\langle \,#1\, \right\rangle}}

\newcommand{\be}{\begin{equation}}
\newcommand{\ee}{\end{equation}}

\newcommand{\orcid}[1]{\href{https://orcid.org/#1}{\includegraphics[width=0.7em]{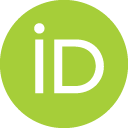}}}

\def\mpcoh{\,h^{-1}{\rm Mpc}}

\def\msolaroh{\,h^{-1}M_\odot}
\def\citejap#1{\citeauthor{#1}\ \citeyear{#1}}
\def\m@th{\mathsurround=0pt }
\def\eqalign#1{\null\,\vcenter{\openup1\jot \m@th
 \ialign{\strut\hfil$\displaystyle{##}$&$\displaystyle{{}##}$\hfil
 \crcr#1\crcr}}\,}

\defcitealias{AlamGAMA}{A21}
\defcitealias{pa20}{PA20}

\begin{document}


\title[VVF and Assembly bias]{Impact of tidal environment on galaxy clustering in GAMA}

\author[Alam et. al.] {\parbox{\textwidth}{
    Shadab Alam\orcid{0000-0002-3757-6359}$^{1,2}$ \thanks{shadab.alam@tifr.res.in},
    Aseem Paranjape\orcid{0000-0001-6832-9273}$^{3}$ \thanks{aseem@iucaa.in},
    John A. Peacock\orcid{0000-0002-1168-8299}$^{2}$\thanks{jap@roe.ac.uk},
    } \vspace*{4pt} \\ 
\vspace{-1.5mm} $^{1}$ Tata Institute of Fundamental Research, Homi Bhabha Road, Mumbai 400005, India \\
\vspace{-1.5mm} $^{2}$Institute for Astronomy, University of Edinburgh, Royal Observatory, Blackford Hill, Edinburgh, EH9 3HJ , UK\\
\vspace{-1.5mm} $^{3}$Inter-University Centre for Astronomy \& Astrophysics, Ganeshkhind, Post Bag 4, Pune 411007, India
}

\date{\today}
\pagerange{\pageref{firstpage}--\pageref{lastpage}}   \pubyear{2018}
\maketitle
\label{firstpage}

\input{tex/abstract}


 \input{tex/intro}

\input{tex/model}

\input{tex/data}

\input{tex/Measurements}

\input{tex/result}

\input{tex/summary}

\section{Data Availability}
All of the observational datasets used in this paper are made available through the GAMA website \url{http://www.gama-survey.org/}. The codes used in this analysis along with instructions will be made available on \url{https://www.tifr.res.in/~shadab.alam/CodeData/}. Some of the $N$-body simulations used in this paper can be accessed through \url{https://www.cosmosim.org/cms/simulations/bolshoi/}.

\input{tex/acknowledgement}


\bibliography{Master_Shadab,morerefs}
\bibliographystyle{mnras}

\label{lastpage}

\end{document}

%% file: tex/abstract.tex
\begin{abstract}
We constrain models of the galaxy distribution in the cosmic web using data from the Galaxy and Mass Assembly (GAMA) survey. 
We model the redshift-space behaviour of the 2-point correlation function (2pcf) and the recently proposed Voronoi volume function (VVF) -- which includes information beyond 2-point statistics. 
We extend the standard halo model using extra satellite degrees of freedom and two assembly bias parameters, $\alpha_{\rm cen}$ and $\alpha_{\rm sat}$, which respectively correlate the occupation numbers of central and satellite galaxies with their host halo's tidal environment.
We measure $\alpha_{\rm sat}=1.44^{+0.25}_{-0.43}$ and $\alpha_{\rm cen}=-0.79^{+0.29}_{-0.11}$ using a combination of 2pcf and VVF measurements, representing a detection of assembly bias at the 3.3$\sigma$ (2.4$\sigma$) significance level for satellite (central) galaxies. This result remains robust to possible anisotropies in the halo-centric distribution of satellites as well as technicalities of estimating the data covariance. 
We show that the growth rate ($f\sigma_8$) deduced using models with assembly bias is about 7\% (i.e. $1.5\sigma$) lower than if assembly bias is ignored. 
When projected onto the $\Omega_m$-$\sigma_8$ plane, the model constraints without assembly bias overlap with Planck expectations, while allowing assembly bias introduces significant tension with Planck, preferring either a lower $\Omega_m$ or a lower $\sigma_8$.  
Finally, we find that the all-galaxy weak lensing signal is unaffected by assembly bias, but the central and satellite sub-populations individually show significantly different signals in the presence of assembly bias. 
Our results illustrate the importance of accurately modelling galaxy formation for cosmological inference from future surveys.
\end{abstract}

%% file: tex/intro.tex
\section{Introduction}
\label{sec:intro}
On very large cosmological scales the light emitted by galaxies provides an observational window into the underlying structure of our Universe. This large-scale structure (LSS) traced by the distribution of galaxies helps measure cosmological parameters to a very high precision \citep[e.g.][]{2020PASJ...72...16H, 2021PhRvD.103h3533A, 2021A&A...646A.140H,2022PhRvD.105b3520A}. 
On large scales, the galaxies tracing the LSS can be treated as point particles sampling the local peaks of the underlying matter distribution, thus tracing the $n$-point functions of the dark matter distribution up to a hierarchy of galaxy bias coefficients \citep{Bardeen1986,Cole1989}.  
In reality, the galaxies themselves are extremely complex objects and there are a number of factors influencing their formation and evolution. One of the important challenges for galaxy formation theory as well as cosmology is to identify the key ingredients influencing the formation and evolution of different types of galaxies, hence allowing a clean interpretation of their $n$-point correlations.

Past major spectroscopic galaxy surveys  (2dFGRS: \citejap{Colless2003}; 6dFGS: \citejap{Jones2009}) measured $\sim 10^5$ redshifts; more recent surveys (SDSS-III: \citejap{Eisenstein2011};  WiggleZ: \citejap{WiggleZ}; DEEP2: \citejap{Deep2013}; VIPERS: \citejap{Garilli2014};  GAMA: \citejap{gama2018}; SDSS-IV: \citejap{2016AJ....151...44D}) have been of the same size or up to a million redshifts; ongoing and future surveys (PFS: \citejap{2014PASJ...66R...1T}; 4MOST: \citejap{2019Msngr.175....3D}; DESI: \citejap{2016arXiv161100036D}) will measure 10-50 million galaxy spectra. These data sets will provide exquisitely precise measurements of the two-point clustering and higher order statistics of the galaxy distribution. The improvement in measurements arises not only from the increased volume but also from improved instruments, leading to accurate measurements of small scale statistics ($\lesssim$\,$1\mpcoh$) for much fainter galaxies. This is especially important as hierarchical structure formation starts from nearly Gaussian density perturbations, making the largest scales fully described by two-point clustering statistics within linear theory, whereas non-linear evolution on small scales affects clustering at all orders, beyond simply two-point clustering. Therefore two important challenges need to be overcome so as to extract maximal information from such data sets: 
(a) Understanding the non-linear impact of gravitational dynamics of dark matter as well as astrophysical processes affecting galaxy formation and evolution at small scales; and
(b) Exploiting the additional information on non-linear evolution that is encoded in beyond two-point statistics.

The first challenge in modelling non-linear scales is that we ideally require a detailed understanding of galaxy formation physics -- and furthermore need to be able to evaluate any such model rapidly and efficiently. Due to the highly non-linear nature of the problem, analytical approaches are limited in scope; it is better to use numerical methods that can accurately track astrophysical hydrodynamics coupled to dark matter through gravity. However, the large dynamic range in the problem requires a number of approximations regarding `sub-grid' physics to be made, and even then the final solutions are still computationally extremely demanding \citep[e.g.][]{2010MNRAS.402.1536S,2014Natur.509..177V,2016MNRAS.463.3948D,2017arXiv170609899T}. An intermediate approach to this problem is to begin by solving the gravity-only evolution numerically using $N$-body simulations; this step can be relatively rapid. One can then build empirical models to connect the gravity-only solution to the galaxy distribution. Again, a number of such empirical methods exist, e.g. semi-analytical galaxy formation models (SAM: \citejap{1991ApJ...379...52W}); sub-halo abundance matching (SHAM; see e.g. \citejap{2004MNRAS.353..189V}; \citejap{2006ApJ...647..201C}); conditional luminosity function (CLF; \citep{2003MNRAS.339.1057Y,2012arXiv1204.0786M}); and halo occupation distribution  \citep[HOD:][]{Benson2000,Seljak2000,Peacock2000,White2001,Berlind2002,Cooray2002,zu2015}. We adopt the empirical HOD approach in this paper as it has been successfully applied to a number of studies while being efficient enough to be computationally feasible for the current analysis. 

A key assumption in the simplest implementation of HOD models is that the distribution of galaxies within a halo only depends on the mass of the host halo. But this may not be valid, and galaxies might preferentially populate certain large-scale halo environments. That would affect summary statistics at all scales, but more strongly on non-linear scales; such environmental effects are known in general as `galaxy assembly bias'. Note that effects of this sort have two distinct aspects: one is the fact that haloes in different large-scale environments will have different assembly histories and hence will exhibit a different large-scale bias even at fixed mass. This effect is known as `halo assembly bias' \citep{2002ApJ...568...52W, 2004MNRAS.350.1385S, 2009MNRAS.399..983A, 2009MNRAS.398.1742H, Shi2017arXiv170704096S}, but it is not in any sense a missing ingredient in standard treatments of halo clustering. As originally shown by \cite{Kaiser1984ApJ...284L...9K}, the bias of objects depends on how rare they are, with the highest peaks being the most strongly biased. Since the highest peaks collapse first, there is obviously a correlation between halo formation redshift and bias, even at a fixed mass scale; but the usual discussion of mass-dependent halo bias automatically averages over this range of bias values. An issue for the HOD approach therefore only arises when the galaxy populations within a halo {\it also\/} depend on formation time. But there is every reason to expect some effect of the latter sort:
galaxy formation processes such as the ability to transport cold gas or supernovae feedback might have different efficiencies at different times. Since haloes in regions of higher large-scale overdensity collapse earlier, there is then indeed scope for galaxy properties to be correlated with large-scale environment \citep{2021MNRAS.508..940M,2022NatAs...6..599D}. These assembly bias effects are important for two reasons. Firstly, measurements of such effects can help us improve our understanding of galaxy formation physics; and secondly, LSS modelling that aims to infer unbiased cosmological parameters risks being systematically in error if we do not properly allow for assembly bias 
in the analysis \citep[e.g.][]{2014MNRAS.443.3044Z}. The aim of this paper is therefore to extend the HOD model by equipping it with degrees of freedom that allow us to produce model galaxy distributions that incorporate assembly bias.

The second challenge of accessing the considerable information encoded in beyond two-point clustering can be addressed using a number of summary statistics. The first natural extension is to use three-point clustering, which has been studied in simulations and observed with high precision by several groups \cite[e.g.][]{2004ApJ...605L..89S,2009MNRAS.399..801G}. But the measurement of higher $n$-point statistics is computationally expensive and very quickly becomes a bottleneck  \citep[although see][]{2015MNRAS.454.4142S}. Other higher order statistics include counts-in-cells \citep[CIC:][]{cbs95,bcgs02,friedrich+18,gruen+18,uhlemann+20,rs21}; the void probability function \citep[VPF:][]{white79a,fry86,ml-r87,eg92,vgph94,sheth96,croton+04,fc13}, $k$-Nearest neighbour statistics \citep[kNN:][]{ba21a,ba21b,fard+21} and the Voronoi volume function \citep[VVF:][henceforth PA20]{pa20}. All of these access the beyond-Gaussian information and in principle depend on all the higher order $n$-point functions. Recently, \citetalias{pa20} have proposed that the VVF is sensitive to a unique combination of cosmological and galaxy formation physics. The VVF is defined as the distribution of cell volumes in the Voronoi tessellation of a set of galaxy positions observed in redshift space. As discussed by \citetalias{pa20}, the VVF is closely connected to the formalism underlying CIC and especially the VPF mentioned above. However, the highly constrained nature of the definition of a Voronoi cell means that  non-linear information is packaged by the VVF in a complex manner that cannot easily be disentangled analytically. Using $N$-body simulations, \citetalias{pa20} showed that the VVF is highly sensitive to the non-linear clustering properties of samples chosen using different selection criteria. Moreover, the measurement of the VVF requires no input in addition to the redshift-space galaxy positions, along with the survey completeness footprint, already routinely used for two-point function analyses. Therefore, we have chosen to use the VVF to include beyond two-point information in our analysis. 

In a recent paper \citep[][henceforth, A21]{AlamGAMA}, some of us have studied in detail the redshift-space clustering of galaxies in the GAMA survey (\citejap{gama2018}). We showed that the large-scale signature of the fluctuation growth rate is fairly robust against the details of small scale galaxy physics. We also showed that the distribution of satellite galaxies in a halo display deviations from the distribution of dark matter in a halo (NFW profile), with the effect being larger for for fainter galaxies. The purpose of the present study is to extend the analysis in \citetalias{AlamGAMA} in three ways. (a) We include VVF measurements from GAMA to include beyond two-point information. (b) We introduce new assembly bias parameters in the model, permitting additional correlation between the tidal anisotropy of the halo (defined later) and the occupation of central and satellite galaxies at fixed halo mass. (c) We use an improved estimate of the covariance matrix based on new simulations, as compared to the conservative covariance used in previous work.  The main focus of this work is to look for robust signatures of beyond halo mass effects in the galaxy distribution and their impact on growth rate measurements.

The paper is organized as follows. We first describe the details of our modelling methodologies including assembly bias parameter to capture new aspects of galaxy-halo connection in section \ref{sec:model}.
We then introduce the GAMA data and how we create magnitude limited samples in section \ref{sec:data}. 
In section \ref{sec:meas}, we describe our measurements from GAMA data and estimates of covariance matrices. Our results are presented in section ~\ref{sec:result} and a summary is in section ~\ref{sec:summary}.


%% file: tex/model.tex
\section{Modelling non-linear scales}
\label{sec:model}

\begin{figure*}
    \centering
    \includegraphics[width=0.95\textwidth]{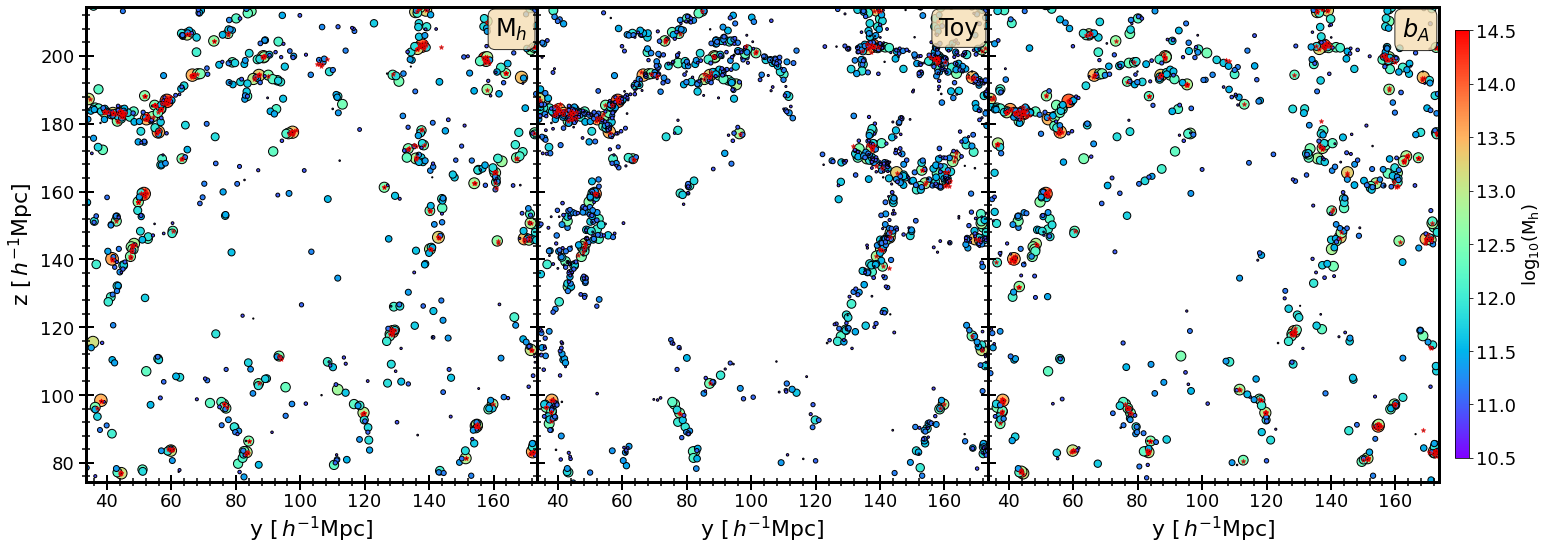}
    \caption{Visualisation of galaxies in a $3 \mpcoh \times 140 \mpcoh \times 140 \mpcoh$ slice, centred at the galaxy with the largest assigned Voronoi volume in the base model. The $y$ and $z$ axes are shown, while we project along the $x$ axis. Circles indicate central galaxies, with colour indicating the host halo mass; the sizes of the circles are also scaled according to the host halo mass. The red stars indicate the satellite galaxies. The left panel is for the mass-only HOD model; the middle panel is for the model with large assembly bias ($\alpha_{\rm cen}=-10$ and $\alpha_{\rm sat}=0$); and the right panel is for the model with a reasonable value of assembly bias that is currently allowed by data ($\alpha_{\rm cen}=-0.71$ and $\alpha_{\rm sat}=1.44$, see section~\ref{subsec:ABHOD}).
   The main feature to see here is that models with assembly bias differ from the base model in that the void regions have fewer galaxies and also that filamentary regions (especially very close to massive objects) are more densely occupied. This is because the assembly bias model with $\alpha_{\rm cen}<0$ makes it less likely for the galaxies to reside in haloes with tidally isotropic environments, while making it more likely for haloes of similar mass in tidally anisotropic filamentary regions to be occupied.}
    \label{fig:xy-void}
\end{figure*}

Perturbation theory describing the growth and evolution of large scale structure (LSS) has been successfully used to model linear and quasi-linear scales \citep{Socco2004,2010PhRvD..82f3522T,2011JCAP...08..012O,ReiWhi11,2012MNRAS.427.2537C,2012JCAP...11..009V}. The only fully reliable way to model non-linear scales, however, is to solve for the exact dark-matter dynamics via $N$-body simulations. We extend the model described in \citetalias{AlamGAMA} in this analysis. Below we summarise the base model briefly and refer readers to \citetalias{AlamGAMA} for more details.

\subsection{Simulation and HOD model}

We assume a flat $\Lambda$CDM cosmology with $\Omega_m=0.27$, $\Omega_b=0.0469$, $h=0.7$, $n_s=0.95$ and $\sigma_8=0.82$. This cosmology is motivated by the fiducial cosmology adopted in the $N$-body simulation \citep[Bolshoi;][]{2011ApJ...740..102K} 
that we employ in our HOD models. 

\begin{table}
    \centering
    \begin{tabular}{l|c|c|c}
    {\bf Simulation} & {\bf MDPL2} & {\bf L300} & {\bf Bolshoi}\\
    \hline
    Cosmology &  Planck & WMAP7 & WMAP7 \\
    Box Size (\!$\mpcoh$) &  1000 & 300 &250 \\
    particle mass ($10 ^{9} \msolaroh$ ) & 1.51 & 1.93 & 0.135\\
    Number of particles & $3840^3$ & $1024^{3}$ & $2048^{3}$\\
    Number of realizations & 1 & 3 & 1 \\
    \end{tabular}
    \caption{Specification of the three simulations used in this analysis. We used MDPL2 and L300 simulations for the covariance matrix estimation and Bolshoi for modelling the properties of the observed galaxy sample.}
    \label{tab:simtab}
\end{table}

Our model predictions are obtained by populating the dark-matter halo catalogue with galaxies using an extended Halo Occupation distribution model. We use a halo catalogue constructed via the ROCKSTAR\footnote{\url{https://bitbucket.org/gfcstanford/rockstar}} halo finder \citep{behroozi13}, which was applied to the snapshot at redshift $z = 0.1$ from the publicly available Bolshoi\footnote{\url{https://www.cosmosim.org/cms/simulations/bolshoi/}} 
simulation \citep{2011ApJ...740..102K}. Bolshoi is a dark matter only $N$-body simulation that uses the Adaptive-Refinement-Tree (ART) code \citep{1997ApJS..111...73K}. The simulation covers a periodic box of side 250$\mpcoh$ with $2048^3$ particles. As stated above, it assumes a flat $\Lambda$CDM cosmology with $\Omega_m=0.27$, $\Omega_b=0.0469$, $h=0.7$, $n_s=0.95$ and $\sigma_8=0.82$. 
Additionally, for covariance matrix calculations we will use two more simulations. The first includes three realisations of a $300\Mpch$ box at $z=0$, simulated with $1024^3$ particles and a flat $\Lambda$CDM cosmology with $\Omega_m=0.276$, $\Omega_b=0.045$, $h=0.7$, $n_s=0.961$ and $\sigma_8=0.811$. Further details of this simulation, which we refer to below as L300, can be found in \citet{ppp19}. The second is from the MultiDark simulation suite, specifically the run MDPL2. This uses a periodic box of 1000$\mpcoh$ on a side with $3840^3$ particles and assumes a flat $\Lambda$CDM cosmology with $\Omega_m=0.31$, $n_s=0.96$, $h=0.67$ and $\sigma_8=0.82$. Both of these simulations were analysed using the ROCKSTAR halo finder. All the simulations are summarised in Table~\ref{tab:simtab}.

We populate the halo catalogues in the simulations using an extended HOD framework as described in \citetalias{AlamGAMA}. This model goes beyond the standard 5-parameter HOD, allowing an additional freedom for  satellites to populate the dark matter haloes. Such non-standard degrees of freedom are particularly important when studying small-scale redshift space clustering, as they allow for more realistic galaxy populations and may introduce degeneracy with the cosmological parameters for high-order statistics. Below, we briefly summarise the details of the HOD model:

\begin{align}
\left\langle N_{\rm cen} \right\rangle_M &= \frac{1}{2}  \mathrm{erfc}\left( \frac{\ln (M_{\rm cut}/M)}{\sqrt{2}\sigma_{\rm M}}\right) \, , \nonumber \\
\left\langle N_{\rm sat}\right\rangle_M &=  N_{\rm cen}(M) \left(  \frac{M-\kappa M_{\rm cut}}{M_1}\right)^\alpha \, 
\label{eqn:HOD}\, ,   
\end{align}
where $\left\langle N_{\rm cen}\right\rangle_M$ gives the occupation probability of central galaxies in a halo of given mass $M$ and average number of satellite galaxies is given by $\left\langle N_{\rm sat}\right\rangle_M$. The central galaxies are placed at the centre of dark matter haloes and given the velocity of dark matter haloes scaled by a free parameter $\gamma_{\rm HV}$. The satellite galaxies are populated using the NFW profile \citep{1996ApJ...462..563N}. The satellite distribution has three additional free parameters $f_c$, $f_{\rm vir}$ and $\gamma_{\rm IHV}$, where $f_c$ scales the concentration of satellites,   $f_{\rm vir}$ scales the maximum radius of the satellites population in unit of $r_{\rm vir}$, and $\gamma_{\rm IHV}$ scales the velocity dispersion of satellite galaxies in unit of the halo velocity dispersion.

\subsection{Assembly bias parameters}
\label{subsec:ABparams}
 One of the main characteristics found in hydrodynamical simulations of galaxy formation is that the galaxy population in a dark matter halo depends on its full evolution history, which is sensitive to the halo environment \citep{2012MNRAS.427.1816G}. There are several secondary variables beyond halo mass shown to be connected to halo assembly bias, such as concentration, tidal environment, local over-density etc. (e.g. \citejap{2004MNRAS.350.1385S}; \citejap{2019MNRAS.489.2977R}). Recently \cite{2018MNRAS.476.3631P} proposed a new secondary variable called tidal anisotropy ($\alpha_{R}$) which is sensitive to halo assembly bias. The tidal anisotropy is defined as
 \begin{equation}
     \alpha_{R} =\sqrt{q^2_R} (1+\delta_R)^{-1},
 \end{equation}
 where $\delta_R$ is the dark matter overdensity and $q^2_R$ is the tidal shear defined on the scale of $R$. Both of these are defined in terms of the eigenvalues ($\lambda_1,\lambda_2,\lambda_3$) of the tidal tensor smoothed on a scale $R$ with a Gaussian filter, using
 \begin{align}
\delta_R &= \lambda_1+\lambda_2+\lambda_3\,,\\
q_R^2 &= \frac12\left[(\lambda_1-\lambda_2)^2+(\lambda_2-\lambda_3)^2+(\lambda_3-\lambda_1)^2\right]\,.
 \end{align}
As demonstrated by \citet{2018MNRAS.476.3631P} and \citet{2019MNRAS.489.2977R}, choosing\footnote{Here $R_{\rm 200b}$ is the halo-centric radius enclosing a density equal to $200$ times the mean background density of the Universe.}  $R=4R_{\rm 200b}/\sqrt{5}$ maximises the correlation between the tidal anisotropy $\alpha_R$ and large-scale halo bias, as well as between $\alpha_R$ and internal halo properties. In fact, as shown by  \citet{2019MNRAS.489.2977R}, $\alpha_R$ thus defined is the primary indicator of halo assembly bias, in that the correlation of large-scale halo bias with a large number of secondary halo properties can be understood as arising from their individual correlations with $\alpha_R$. We use this definition of tidal anisotropy and refer the reader to \citet{2018MNRAS.476.3631P} for full details concerning the measurement of $\alpha_R$ using simulated haloes.

The tidal anisotropy (using a constant smoothing scale) has been measured in data from the Sloan Digital Sky Survey (SDSS) and shown to generate a large variation in bias, independent of local over-density \citep{2018MNRAS.476.5442P,azpm19}. Note that perturbation theories \citep{2012PhRvD..85h3509C,2012PhRvD..86h3540B} as well as the effective field theory of large scale structure \citep[e.g.,][]{2015JCAP...11..007S}  also include such tidal bias terms. But these are proportional to the tidal shear: for a Gaussian random field, this is independent of over-density but for the non-linear dark matter density field it shows strong correlations with over-density. This is one of the key reasons for us to use tidal anisotropy, which is approximately independent of the over-density field (the primary driver of gravitational growth).
As mentioned above, the tidal anisotropy is a strong indicator of halo assembly bias. To investigate whether galaxies inherit any of this environmental dependence, we introduce two new HOD parameters ($\alpha_{\rm cen}$ and $\alpha_{\rm sat}$), which modify the occupation of dark matter haloes depending on tidal anisotropy. Following \citet{xzc21}, our parametrization of assembly bias is as follows:
\begin{align}
    \log M_{\rm cut} (\alpha_R) &= \log M^0_{\rm cut} + \alpha_{\rm cen} \times [\alpha^{\rm rank}_R -0.5] \label{eq:assembly_cen}\\ 
    \log M_{\rm 1} (\alpha_R) &= \log M^0_{\rm 1} + \alpha_{\rm sat} \times [\alpha^{\rm rank}_R -0.5]\, ,
    \label{eq:assembly}
\end{align}
where $\alpha^{\rm rank}_R$ is the rank of tidal anisotropy in fine bins of halo mass divided by the total number of haloes in the respective mass bins, so that $\alpha^{\rm rank}_R$ varies between 0 and 1. 
The above equations modify the $M_{\rm cut}$ and $M_{\rm 1}$ HOD parameters as a function of tidal environment. For a positive value of the parameters $\alpha_{\rm cen}$ and $\alpha_{\rm sat}$, this will mean that the haloes with more tidally anisotropic environment will have a higher mass limit for assigning central galaxies and fewer satellites compared to haloes in tidally isotropic environment. Using the ranks of tidal anisotropy rather than the actual values makes us less sensitive to the exact definition and allow us to probe the effect across the whole mass range with a simple mass independent parameterization. It is also important to emphasise that the rank of tidal anisotropy is assigned in narrow mass bins, thus ensuring that any non-zero assembly bias signature is distinct from a model allowing a more complicated mass dependence.

We show a small sub-volume of the simulated galaxy catalogue in Figure~\ref{fig:xy-void} in order to build a more intuitive feeling for the assembly bias model. The panels show (left to right) a mass only HOD model; then a model with large assembly bias; and finally a  model with values of the assembly bias parameters that are consistent with current data.
We see that including assembly bias in this case causes void regions to have fewer galaxies, while filamentary regions (especially very close to massive objects) become more densely occupied.  This is mainly because we are assuming negative values of $\alpha_{\rm cen}$: the tidally isotropic regions around void centres will then show a larger cut-off mass for assigning central galaxies. As a result, the probability of having central galaxies in low mass haloes is reduced, and since these haloes predominate in voids the centre of voids will thus have fewer central galaxies.

%% file: tex/data.tex
\section{Data}
\label{sec:data}

Galaxy and Mass Assembly (GAMA) is a flux limited spectroscopic survey described in \cite{2015MNRAS.452.2087L} and \cite{2018MNRAS.474.3875B}. GAMA provides the redshift measurement of approximately 300,000 galaxies selected from SDSS imaging \citep{2009ApJS..182..543A} with target selection defined in \cite{2010MNRAS.404...86B}. It covers a total sky area of $230$ deg$^2$ with 98\% redshift completeness down to $r$-band Petrosian magnitude 19.8. We use three $5\times12\,{\rm deg}^2$ GAMA equatorial regions G09, G12 and G15, centred on 9h, 12h and 14.5h in right ascension. We create a magnitude limited sample with $M_r<-19$ after applying the $k$-correction and evolution correction \citep{Blanton2003, 2012MNRAS.420.1239L, 2015MNRAS.451.1540L}. We use the DR3 data release \citep{gamaDR3} in this analysis. We refer the reader to \citetalias{AlamGAMA} for more details about the sample selection.

%% file: tex/Measurements.tex
\section{Measurements}
\label{sec:meas}

\begin{figure}
    \centering
    \includegraphics[width=0.47\textwidth]{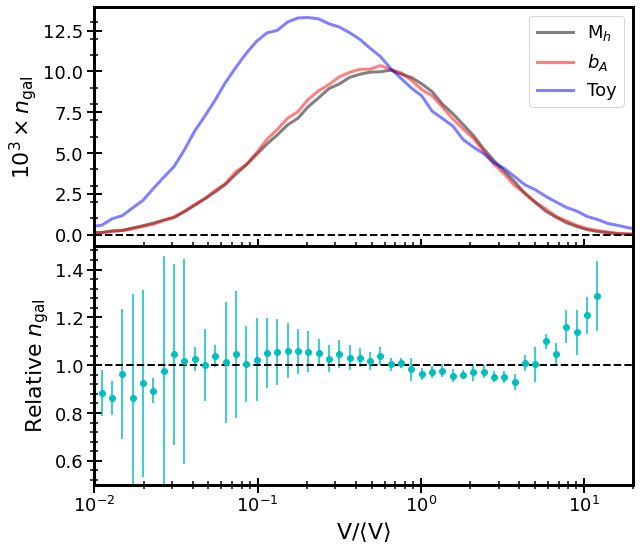}
    \caption{ Distribution of relative volume for galaxies, $V(g)/\left< V \right>$, from three different simulated galaxy catalogues. The black line in the top panel is for a galaxy catalogue with a mass only HOD model; the blue line is for a model with large assembly bias ($\alpha_{\rm cen}=-10$ and $\alpha_{\rm sat}=0$);  and the red line is for a model that depends on both mass and tidal anisotropy ($\alpha_{\rm cen}=-0.71$ and $\alpha_{\rm sat}=1.44$ from Table~\ref{tab:hodpar}). The bottom panel shows the ratio of red and black lines, with an error estimated from jackknife sampling, reflecting the noise for a survey of volume $[250 \mpcoh]^3$.}
    \label{fig:vvf-hist}
\end{figure}

\begin{figure}
    \centering
    \includegraphics[width=0.47\textwidth]{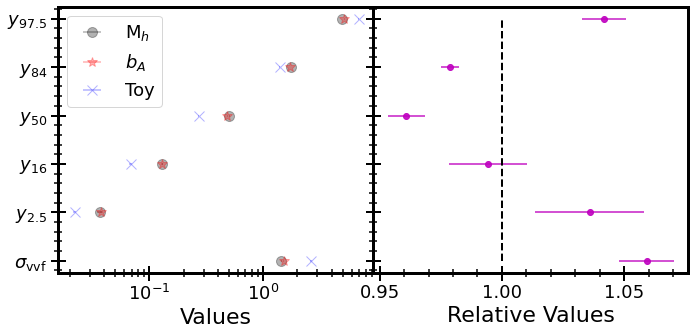}
    \caption{ Comparison of VVF percentile and width for the three models, one with a mass only HOD (shown in black) and an alternative in which the occupation also depends on tidal anisotropy (shown in red). The blue points show the model with large assembly bias ($\alpha_{\rm cen}=-10$ and $\alpha_{\rm sat}=0$). The right panel shows the ratio of the two models (red and black) with errors from jackknife resampling for a survey of volume $[250 \mpcoh]^3$ }
    \label{fig:vvf-percentile}
\end{figure}

\begin{figure*}
    \centering
    \includegraphics[width=0.98\textwidth]{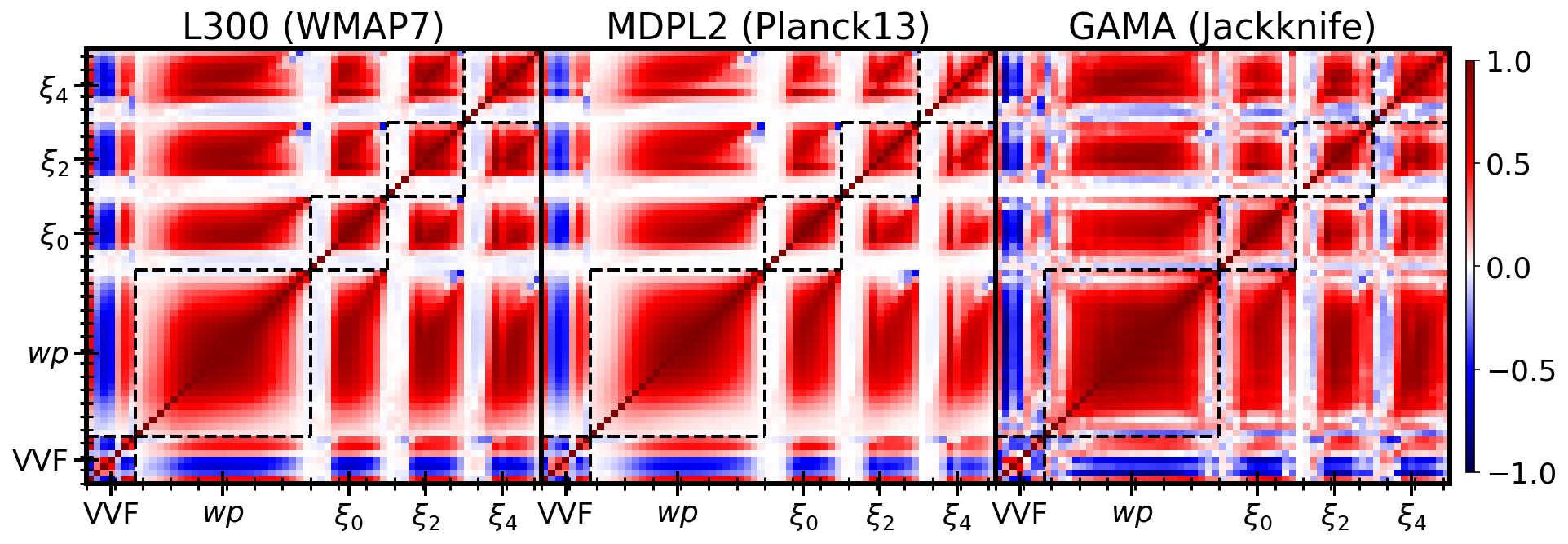}
    \caption{Correlation matrices estimated using three different methods. The left and middle panels shows the correlation matrix estimated using the mean of jackknifes of mocks for L300 and MDPL2 simulation and the rightmost panel shows the one estimated using the jackknife of data. The different diagonal blocks are indicated by the black dashed lines. The first blocks shows the VVF in order of $\sigma_{\rm VVF}$, $y_{2.5}$, $y_{16}$, $y_{50}$, $y_{84}$, $y_{97.5}$ and $\bar{n}$. The second block shows the projected correlation function $w_p$ and the next three blocks show the multipoles  (i.e $\xi_0$, $\xi_2$ and $\xi_4$).
    }
    \label{fig:corr-vvf}
\end{figure*}

\begin{figure*}
    \centering
    \includegraphics[width=0.98\textwidth]{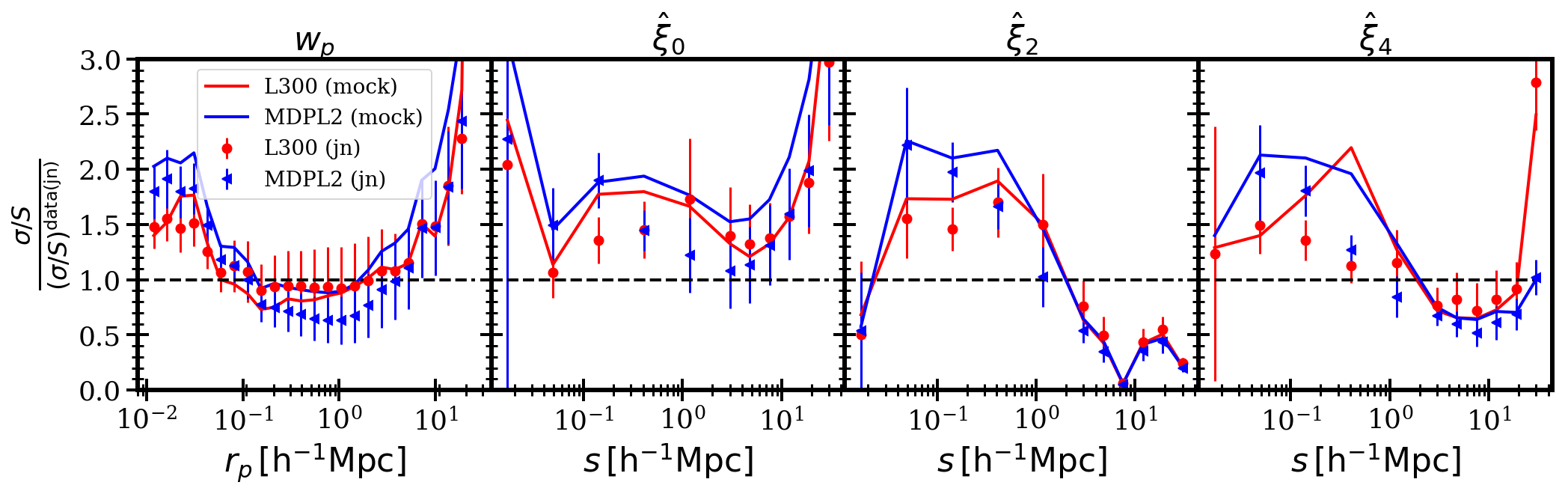}
    \caption{Ratio of signal to noise estimated from mocks to the one estimated from jackknife resampling of the data. The four panels from left to right represent $w_p$, $\xi_0$, $\xi_2$ and $\xi_4$. The solid lines shows the estimates from the variance of the mocks and the points with errors show the estimates from mean of jackknife resamples from the mocks, with error bars indicating the variance of the jackknifes between mocks. The red colour is for the mocks using the L300 simulation and the blue corresponds to the one using the MDPL2 simulations.}
    \label{fig:SNR-wpxi}
\end{figure*}

\begin{figure}
    \centering
    \includegraphics[width=0.47\textwidth]{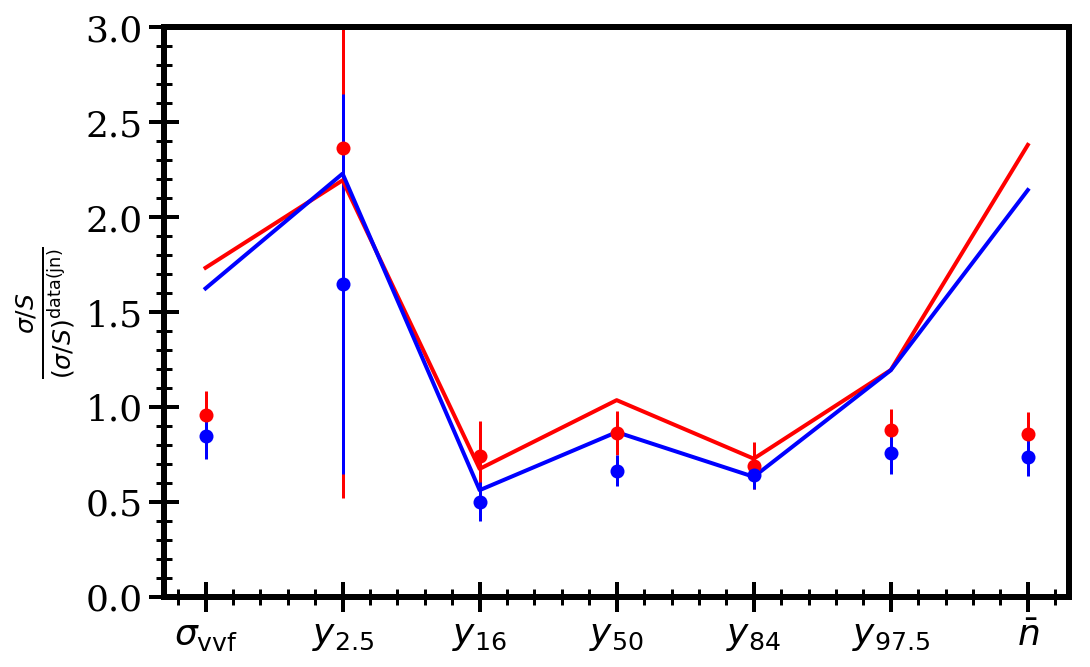}
    \caption{The same as Figure~\ref{fig:SNR-wpxi}, but for VVF-related statistics.}
    \label{fig:SNR-vvf}
\end{figure}

\subsection{Redshift space galaxy clustering}
\label{subsubsec:clustering}

We measure the galaxy auto-correlation function using the minimum variance Landay-Szalay estimator \citep{LandySzalay93}. We then project the two-dimensional galaxy correlation into two different statistics: the projected correlation function, $w_{\rm p}(r_\bot)$, and multipoles ($\xi_{\ell=0,2,4}$). The projected correlation function is obtained by integrating the correlation function along the line-of-sight at fixed perpendicular separation (see equation 19 in \citetalias{AlamGAMA}). For the projected correlation function we adopt a maximum integration limit of $|\pi_{\rm max}|=40 \mpcoh$. We note that in the limit of $\pi_{\rm max}\to\infty$ the projected correlation function becomes independent of redshift space distortions (RSD), which is attractive from a modelling point of view. Since we use a finite $\pi_{\rm max}$, limited by the considerations of signal-to-noise and computing time needed for the model evaluation, some RSD remain in the projected correlation function once we reach $r_p$ comparable to $\pi_{\rm max}$, but our model and covariance matrix both account for the finite integration length. 
The multipoles are obtained by integrating over the angle from the line-of-sight, weighted with the appropriate Legendre polynomial (see equation 20 in \citetalias{AlamGAMA}).

The projected correlation function was measured in 25 log-spaced bins between $0.01\mpcoh$ -- $30\mpcoh$. 
The multipoles were measured with two components, first the small scale component was measured as three wedges with five log-spaced bins between $0.01\mpcoh$ -- $2\mpcoh$ and the larger scales were measured with six log-space bins centred at $3\mpcoh$ -- $30\mpcoh$.

\subsection{Voronoi Volume Function (VVF)}
\label{subsubsec:vvf}
The measurement of the VVF requires us to divide the volume covered by the observed sample of galaxies using a Voronoi tessellation. In principle one could do this using geometric algorithms, but in real surveys one also needs to account for masked regions as well as incomplete regions. These effects are usually accounted for by first generating spatially uniform random points and then applying the survey mask and incompleteness to this original uniform sample. Therefore, another way to measure the VVF is simply by counting the number of randoms associated with each galaxy (\citealp{azpm19}; \citetalias{pa20}). This is defined by linking each random point to its nearest galaxy, then shifting focus to the galaxies and counting the number of randoms $\nu_{\rm ran}(g)$ linked to each galaxy $g$. We can estimate the volume $V(g)$ of the Voronoi cell associated with each galaxy as $V(g)=V_{\rm tot}\nu_{\rm ran}(g)/N_{\rm ran}$. Where $N_{\rm ran}$ and $V_{\rm tot}$ are the total number of randoms and total volume of the region we are analysing. The volume associated with each galaxy scaled by the average volume can be written as $V(g)/\avg{V}= \nu_{\rm ran}(g)\,N_{\rm gal}/N_{\rm ran}$, where $N_{\rm gal}$ is the total number of galaxies in the considered volume. This raises the question of the optimum number of random points one should use in performing such measurements. Using too many random points wastes computing time, but having too few randoms will introduce large errors into our estimates of the VVF, especially of its low percentiles, which one needs to avoid. \citetalias{pa20} reported the results of convergence tests for the ratio $N_{\rm ran}/N_{\rm gal}$. Based on these and our own tests, in this work we set $N_{\rm ran}/N_{\rm gal}=500$. The volume assigned to galaxies at the boundary of masks and edge of survey will be incorrectly estimated. Appendix B2 of \citetalias{pa20} quantified this effect and showed that even in the presence of $10\%$ masked area the effect of such errors on the overall VVF percentile is negligible at the current precision. Such effects, however, will need to be dealt with more exactly for future larger samples such as DESI, PFS and 4MOST.

The properties of the VVF for halo populations and abundance matched galaxy catalogues is discussed in detail in our earlier work, \citetalias{pa20}. The properties of the VVF as a function of cosmology and HOD model are discussed in Liya et. al. in prep. . Figure~\ref{fig:vvf-hist} shows the distribution of the VVF for different galaxy populations. 
The black line represents the base model with no assembly bias. The red line is for a model with a degree of assembly bias currently allowed by GAMA, and the blue line is a toy model with unrealistically large assembly bias.
These black and red populations are generated such that the two point function as well as the number density are consistent within the GAMA errors in each case, but one uses a mass-only HOD model whereas the other includes additional parameters for assembly bias, as described in section~\ref{sec:model}, with $\alpha_{\rm cen}=-0.71$ and $\alpha_{\rm sat}=1.44$ from Table~\ref{tab:hodpar}. Note that the overall VVF distribution is very similar in these two models constrained by the number density and redshift space correlation function. In detail, though, we find that the bottom plot shows significant differences for galaxies having large Voronoi volumes (corresponding to underdense, void-like environments). The model with assembly bias produces more galaxies with these larger volumes, so that voids will tend to be emptier. 

The distribution of the VVF is difficult to use in a likelihood analysis, as accounting for cross-covariance with two-point statistics becomes a computational challenge. Therefore, we reduce the entire distribution to six numbers giving the scaled Voronoi volume ($V(g)/\avg{V}$) at the $2.5^{\rm th}$, $16^{\rm th}$, $50^{\rm th}$, $84^{\rm th}$ and $97.5^{\rm th}$ percentiles of the VVF  (with the $Q^{\rm th}$ percentile denoted `${\rm y}_{Q}$' and the collection denoted `${\rm VVF}_{\rm p}$'), as well as its standard deviation (denoted by `$\sigma_{\rm VVF}$'). Broadly speaking, lower percentiles such as ${\rm y}_{2.5}$ and ${\rm y}_{16}$ probe small-scale, highly clustered regions (e.g., satellites in groups or clusters) while higher percentiles such as ${\rm y}_{97.5}$ probe large, under-dense regions such as voids. 

Figure~\ref{fig:vvf-percentile} shows the VVF percentile and standard deviation for the same models as in Figure~\ref{fig:vvf-hist}. Note that the model parameters are chosen such that the level of assembly bias is allowed by the current data to emphasise that we are looking for small and subtle effects. To highlight the differences between the (very precise) VVF measurements in the two models, the right panel shows the ratio of measurements in the two models along with jackknife errors. We note that the model with assembly bias gives larger values for $y_{2.5}$ and $y_{97.5}$ compared to the case without assembly bias. Effectively the VVF distribution is skewed towards higher $V/\avg{V}$, which also raises $\sigma_{\rm VVF}$. We also want to highlight that these subtle differences between the VVF distributions of the two models are easily captured in the percentiles and width, and hence these constitute an effective compression of the full VVF distribution.
Note that the error bars in Figures~\ref{fig:vvf-hist} and~\ref{fig:vvf-percentile} represent the noise in an ideal survey with volume $[250 \mpcoh]^3$.

We will exclude the ${\rm y}_{16}$ percentile from our current analysis, as we have found that it is highly correlated with ${\rm y}_{84}$ and ${\rm y}_{2.5}$, making the inverse of the covariance matrix (discussed below) ill-defined.

\subsection{Covariance matrix}
\label{sec:cov}
The covariance matrix quantifying the correlation between various observed statistics is a crucial component in the measurement of various physical parameters. It is also crucial for detecting any beyond halo mass effects in the observed data. To estimate the covariance matrices we first produce two different sets of mock catalogues mimicking the GAMA  survey geometry. We then measure the jackknife covariance matrix from the observed GAMA data and compare it to the mock and jackknife covariance matrices estimated from the two sets of covariance mocks. ~\cite{2022SCPMA..6509811S} have recently shown that cosmic variance is important and thus that the jackknife covariance might not be sufficiently precise, providing further motivation for the tests in this section.

To generate mock galaxy catalogues with the GAMA survey geometry, we first take the periodic box simulations and remap them to a cuboid following the algorithm proposed in \cite{2010ApJS..190..311C}. The transformations are done such that the $z$-axis corresponds to the size of the survey along the light-of-sight for the $M_r<-19$ sample. The ratio of side lengths along the $x$ and $y$ axes is kept as close to the actual value for the GAMA fields as possible. We then divide the cuboid into separate pieces, each presenting one realisation of one of the GAMA field. The fields are also aligned in a close packing by alternating the line-of-sight from the positive to the negative $z$-axis in consecutive regions. Since the three GAMA fields are equivalent we first count the total number of fields that can be generated from the cuboids and use consecutive regions to generate the same field. This is to avoid generating the different fields by using neighbouring regions of the same realisation and thus introducing a spurious larger scale correlation between the mock fields. We use two sets of mocks that were created via this procedure: the first set consists of 50 realisations using the L300 simulation (with a WMAP7 cosmology); and the second set consists of 267 realisations using the MDPL2 simulation (with a Planck cosmology). We populate the halo catalogue in each simulation using the best fit HOD model obtained in \citetalias{AlamGAMA} by fitting the clustering statistics.

Figure~\ref{fig:corr-vvf} shows the correlation matrices obtained using the two sets of mock catalogues L300 and MDPL2, as well as the jackknife covariance estimated from data. The two mock covariance matrices are the mean of the jackknife covariance matrices from individual realisations. The jackknife covariance estimated from the observed data is noisier compared to the mock covariances due to the limited number of jackknife regions. The mock covariances using L300 and MDPL2 are fairly consistent with each other, despite using completely different inherent simulations and slightly different cosmological parameters. We can therefore be reassured that the covariance matrices are robust and do not require an unrealistically precise knowledge of the true cosmological model.

Figure~\ref{fig:SNR-wpxi} shows the ratio of signal-to-noise estimated using mocks to the one estimated using data jackknife, for our various clustering statistics. The columns from left to right show $w_p$, $\xi_0$, $\xi_2$ and, $\xi_4$. We use the ratio of signal-to-noise in this illustration as this will be insensitive to any small differences in the signal itself. We find that the signal to noise estimated from the mean jackknife of the mocks is consistent with direct mock-based estimates. However, both the mock based estimates show larger noise for $w_p$ and the monopole at large scales, compared to the one estimated using the jackknife of the data. Figure~\ref{fig:SNR-vvf} again shows a ratio of signal-to-noise as in Figure~\ref{fig:SNR-wpxi}, but now for the VVF percentile and galaxy number density. We note that in this plot the signal-to-noise ratio estimated by the variance between mocks disagrees with the one estimated via the mean jackknife. We find that quantities such as number density and $\sigma_{\rm VVF}$ show a larger dispersion between the mocks compared to the noise estimated from jackknife sampling. This might be because the number density in a small volume such as GAMA is sensitive to the presence of any massive structure which can appear and disappear among mock realisations. Since these massive structure are rare in small volumes, they possibly violate the independence assumption for jackknife regions. We therefore felt that it was important to make sure that our covariances reflected this larger variance derived from the dispersion between different mocks. But at the same time, it is also important to keep the noise in the covariance as low as possible. We see no obvious differences between the {\it correlation\/} matrix (\smash{$R_{ij}=C_{ij}/[C_{ii} C_{jj}]^{1/2}$}) estimated using different simulations or different methods (e.g. with or without jackknifes). Our covariance matrix is given by the following equation:

\begin{align}
C^{\rm final}_{ij} & = R^{\rm MDPL2}_{ij} \times\sigma_{i} \times\sigma_{j}; \\
 \label{eq:covmat}
\sigma_{i} &= \sqrt{C^{\rm MDPL2}_{ii}} \frac{S_{\rm data}}{S_{\rm MDPL2}} \, .\nonumber
\end{align}

We therefore decided to adopt the correlation matrix estimated from the mean of the jackknifes for MDPL2 mocks ($R^{\rm MDPL2}_{ij}$), since this appears to have the lowest noise. Our full covariance matrix ($C^{\rm final}_{ij}$) is then derived by scaling this correlation matrix using the diagonal terms estimated from the variance between mocks ($C^{\rm MDPL2}_{ii}$) -- with a final additional scaling by the ratio of the signal between data ($S_{\rm data}$) and the MDPL2 mocks ($S_{\rm MDPL2}$), to account for any differences in overall fluctuation amplitude.

\begin{table}
    \centering
    \begin{tabular}{|p{1.2cm}|p{4.5cm}|p{1.5cm}|} \hline\hline
    Parameters &  Description & prior \\ [0.5ex] \hline\hline 
    $M_{\rm cut}$ & Halo mass at which probability of having central galaxy is 0.5.  & $10^{11}$--$10^{15}$ \\ 
    $\sigma_{M}$ & scatter in the halo mass to model the given absolute magnitude limited sample. This should be related to scatter in halo mass and absolute magnitude of galaxies. & 0--8\\
    $\kappa$ & This determines the mass at which haloes have no satellite galaxies in units of $M_{\rm cut}$ & 0--3\\
    $M_{1}$ & This determines the scaling of number of satellite galaxies with halo mass. &  $10^{11}$--$10^{15}$\\
    $\alpha$ & The power law index of number of satellite as the function of halo mass. & 0--3 \\ \hdashline
    $f_c$ & The distribution of galaxies might follow a different concentration than dark matter itself. This parameter scales the concentration of the dark matter halo to determined the concentration of the satellite galaxies by scaling the scale radius $R_s$ of the halo. & $10^{-3}$--5 \\
    $\gamma_{\rm HV}$ & This parameters scales the inter-halo velocity to allow an additional degree of freedom as the growth rate of structure. & 0--3\\
    $\gamma_{\rm IHV}$ & This scales the velocity dispersion of the dark matter halo in order to allow the satellite galaxy velocity distribution to be different from dark matter. & 0--3\\ 
    $f_{\rm vir}$ & This scales the maximum distance up to which satellite galaxies are distributed in unit of virial radius of the halo. A corresponding velocity dispersion is also estimated based on the according to the distance. & 0.1--5\\ \hdashline
    $\alpha_{\rm cen}$ & Assembly bias for central galaxy, correlates the occupation probability with host halo's tidal anisotropy. & $-2.0$--2.0 \\
    $\alpha_{\rm sat}$ & Assembly bias for satellite galaxy, correlates the number of satellites with host halo's tidal anisotropy. & $-2.0$--2.0\\ \hline\bottomrule
    \end{tabular}
    \caption{Description of model parameters and their prior range.}
    \label{tab:parlist}
\end{table}

\subsection{Analysis methods}
\label{sec:analysis}

We start with the ROCKSTAR halo catalogue of the Bolshoi N-body simulation. We first measure the tidal anisotropy for each host halo and then obtain the rank of tidal anisotropy in fine mass bins as described in section ~\ref{subsec:ABparams}. The full parameter space used in this analysis is given in Table \ref{tab:parlist}. We sample this parameter space via an MCMC analysis using {\tt emcee}. We first assign the number of central and satellite galaxies using equation \ref{eqn:HOD} with $M_{\rm cut}$ and $M_1$ redefined in equations~\ref{eq:assembly_cen} \& \ref{eq:assembly} to include the dependence on tidal anisotropy ranks. We use the centre of the halo as the position for the central galaxy, with velocity given by the halo's core velocity scaled by the parameter $\gamma_{\rm HV}$. The satellite galaxies are assumed to follow the NFW distribution given by host halo parameters but the concentration of the distribution is re-scaled by the parameter $f_c$. The velocities are sampled from the Gaussian distribution with mean set to the halo velocity and dispersion given by host halo velocity dispersion multiplied by the parameter $\gamma_{\rm IHV}$. The redshift-space position of each galaxy is obtain by adopting the plane parallel approximation with the $z$-axis of the periodic box being the line-of-sight . 

We now treat this dataset as an observed catalogue and measure all of our observables, namely number density, $w_p$, $\hat{\xi}_{0,2,4}$and VVF. These results are then used as our model prediction. We then calculate the likelihood by estimating the $\chi^2$ which results in the posterior distribution of our parameters given the data and covariance matrix. Our main analysis shows results for four cases as follows:
\begin{enumerate}
    \item $w_p$ + $\xi_{0,2,4}$:  In this case we run chains excluding VVF measurements and fix the tidal anisotropy parameters to zero ($\alpha_{\rm cen}=\alpha_{\rm sat}=0$). Hence, no assembly bias is used in this model.
    \item $w_p$ + $\xi_{0,2,4}$ + ($\alpha_{\rm cen},\alpha_{\rm sat}$): Same as the first case, but allowing both tidal anisotropy parameters to be free and hence allowing for assembly bias in the model.
    \item $w_p$ + $\xi_{0,2,4}$ + {\rm VVF}: Same as the first case, but including VVF in the measurement. In this case we use the chains from the first case and importance sample them due to the expensive VVF calculation.
    \item $w_p$ + $\xi_{0,2,4}$ + {\rm VVF} + ($\alpha_{\rm cen},\alpha_{\rm sat}$): We use all the measurements and allow all parameters including assembly bias parameters to be free. In this case we use the chains from the second case and importance sample them due to the expensive VVF calculation.
\end{enumerate}

%% file: tex/result.tex
\section{Results}
\label{sec:result}

\begin{figure}
    \centering
    \includegraphics[width=0.48\textwidth]{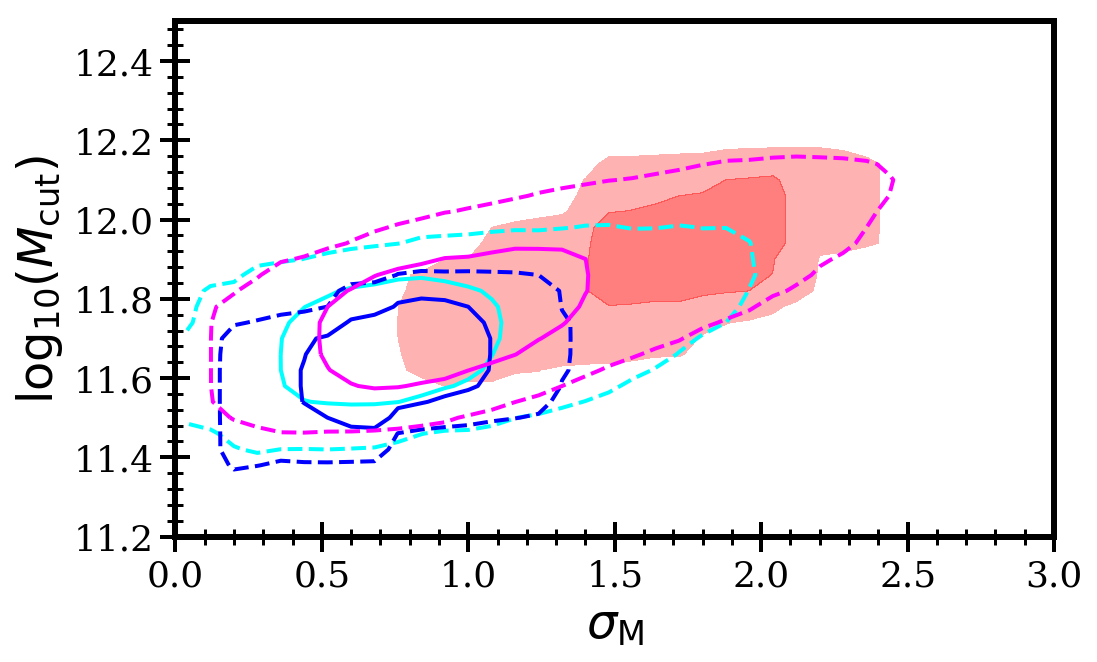}
    \includegraphics[width=0.48\textwidth]{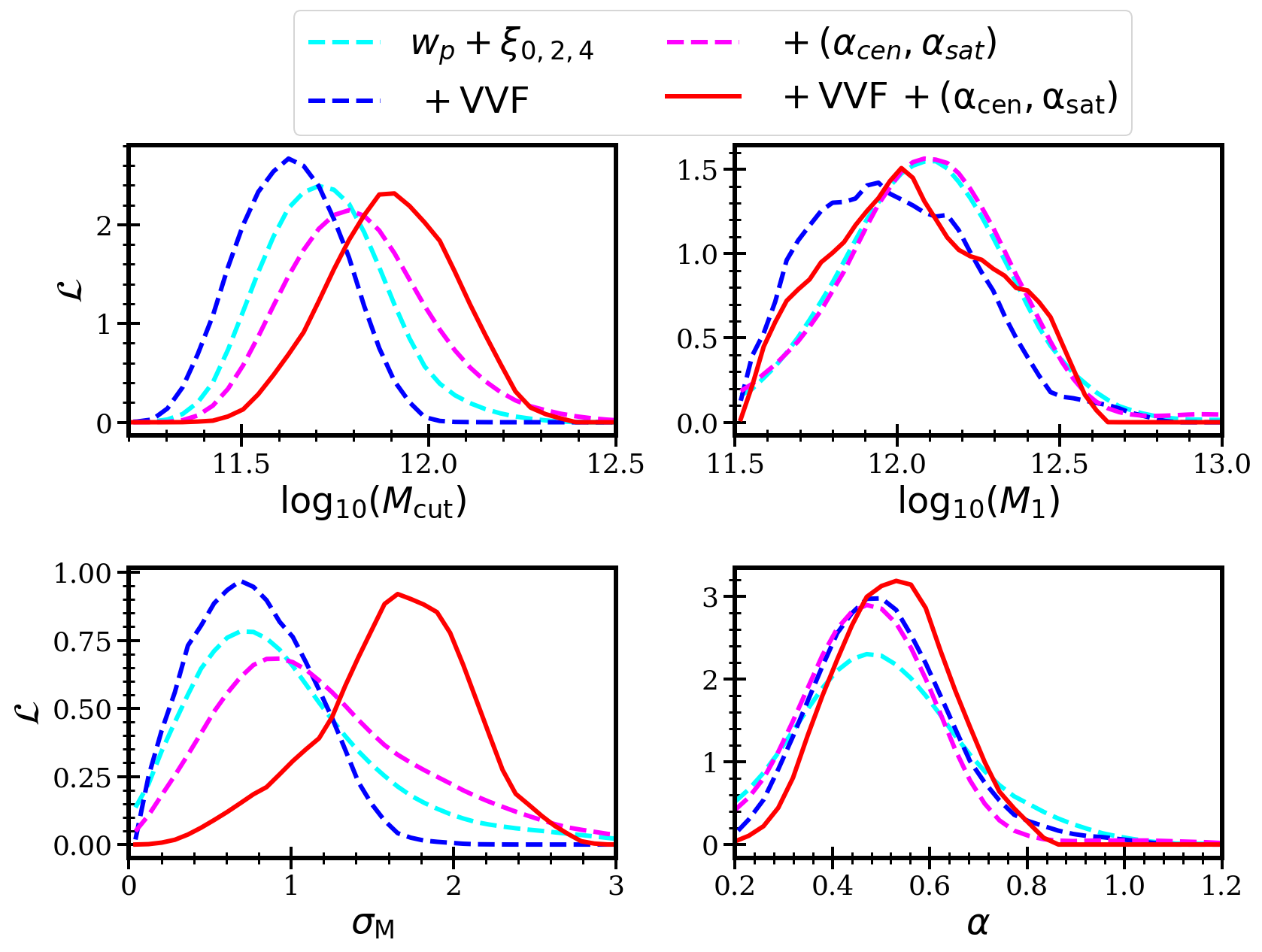}
    \caption{The two-dimensional and one-dimensional constraints on the base HOD parameters for the $M_r<-19$  galaxy sample in GAMA. The 
    The solid and dashed contours of given colour represent the 1$\sigma$ and 2$\sigma$ confidence limits respectively. 
    The magenta and cyan colours indicate the fit to only clustering (i.e. $w_p$ and $\xi_{0,2,4}$) with and without assembly bias parameters respectively. The red and blue colours indicate the fit to clustering and VVF statistics (i.e. $w_p$, $\xi_{0,2,4}$ and VVF percentiles) with and without assembly bias parameters, respectively. In the two-dimensional contours we only show the $M_{\rm cut}-\sigma_M$ plane to highlight the main effect; other parameters are consistent with the results in Figure 7 of \citetalias{AlamGAMA}}
    \label{fig:HODbase}
\end{figure}

\begin{figure*}
    \centering
    \includegraphics[width=1.0\textwidth]{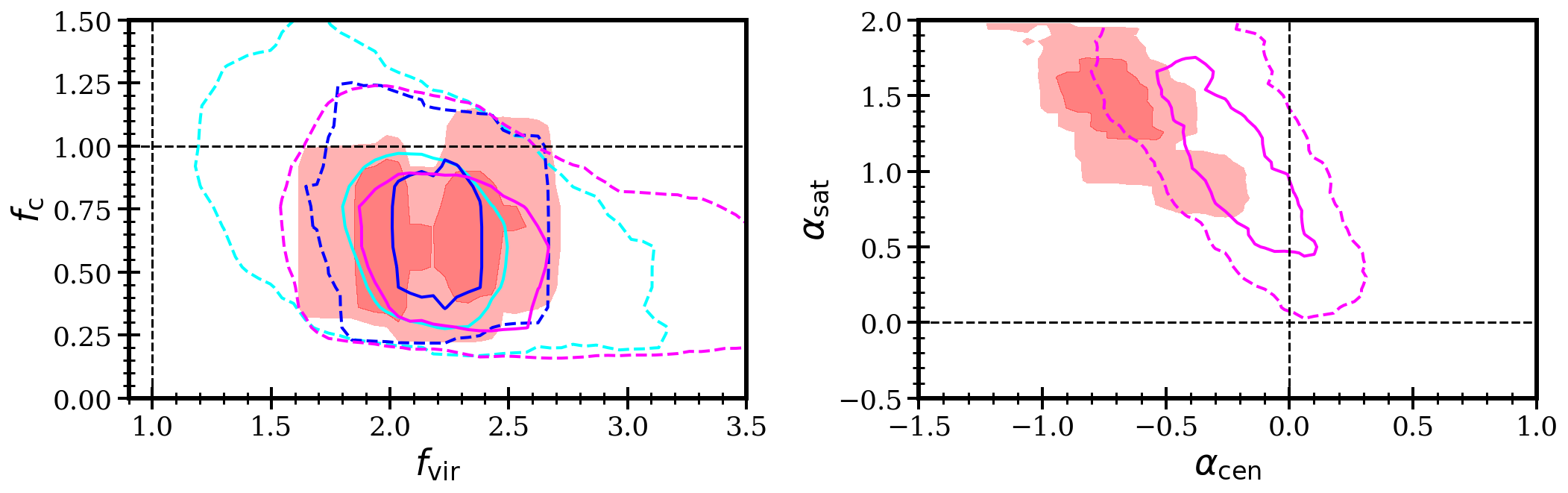}
    \includegraphics[width=1.0\textwidth]{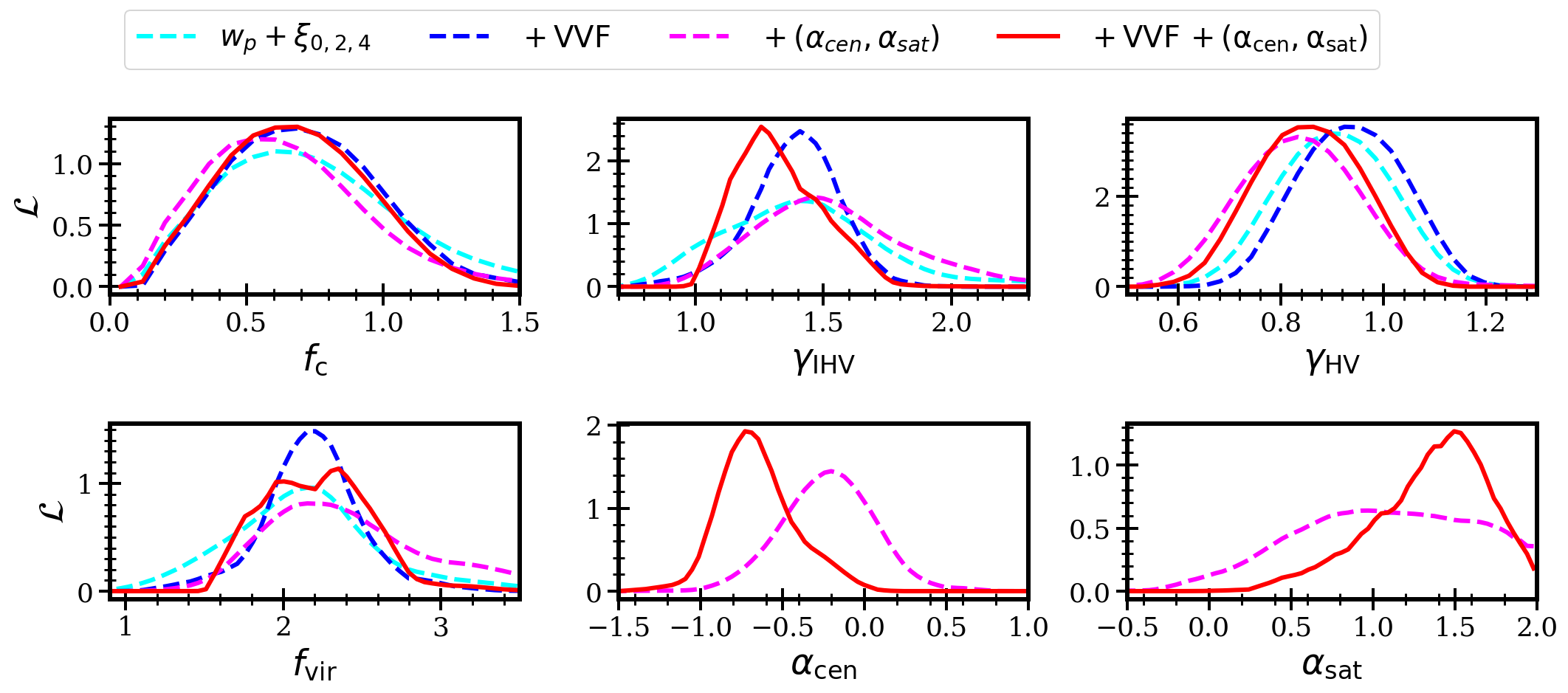}
    \caption{The same as Figure~\ref{fig:HODbase}, but going beyond the base HOD parameters. The one-dimensional likelihoods for the two assembly bias parameters ($\alpha_{\rm cen}$ and $\alpha_{\rm sat}$) do not show much preference for non-zero values in clustering only fits (magenta dashed line), but when including VVF in the fit $\alpha_{\rm sat}$ shows a strong preference for non-zero positive values and $\alpha_{\rm cen}$ for non-zero negative values.}
    \label{fig:HODext}
\end{figure*}

\begin{figure}
    \centering
    \includegraphics[width=0.47\textwidth]{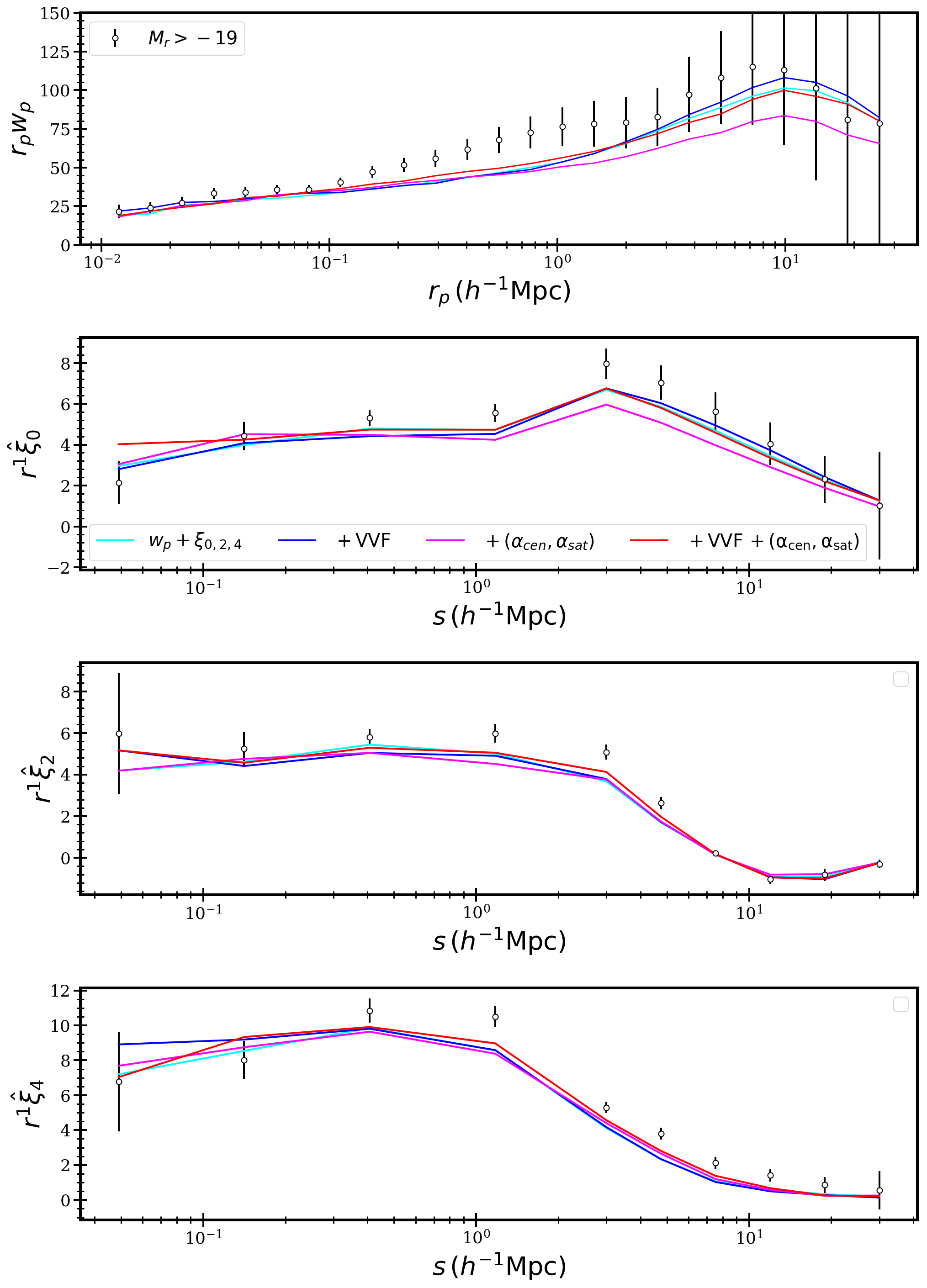}
    \includegraphics[width=0.47\textwidth]{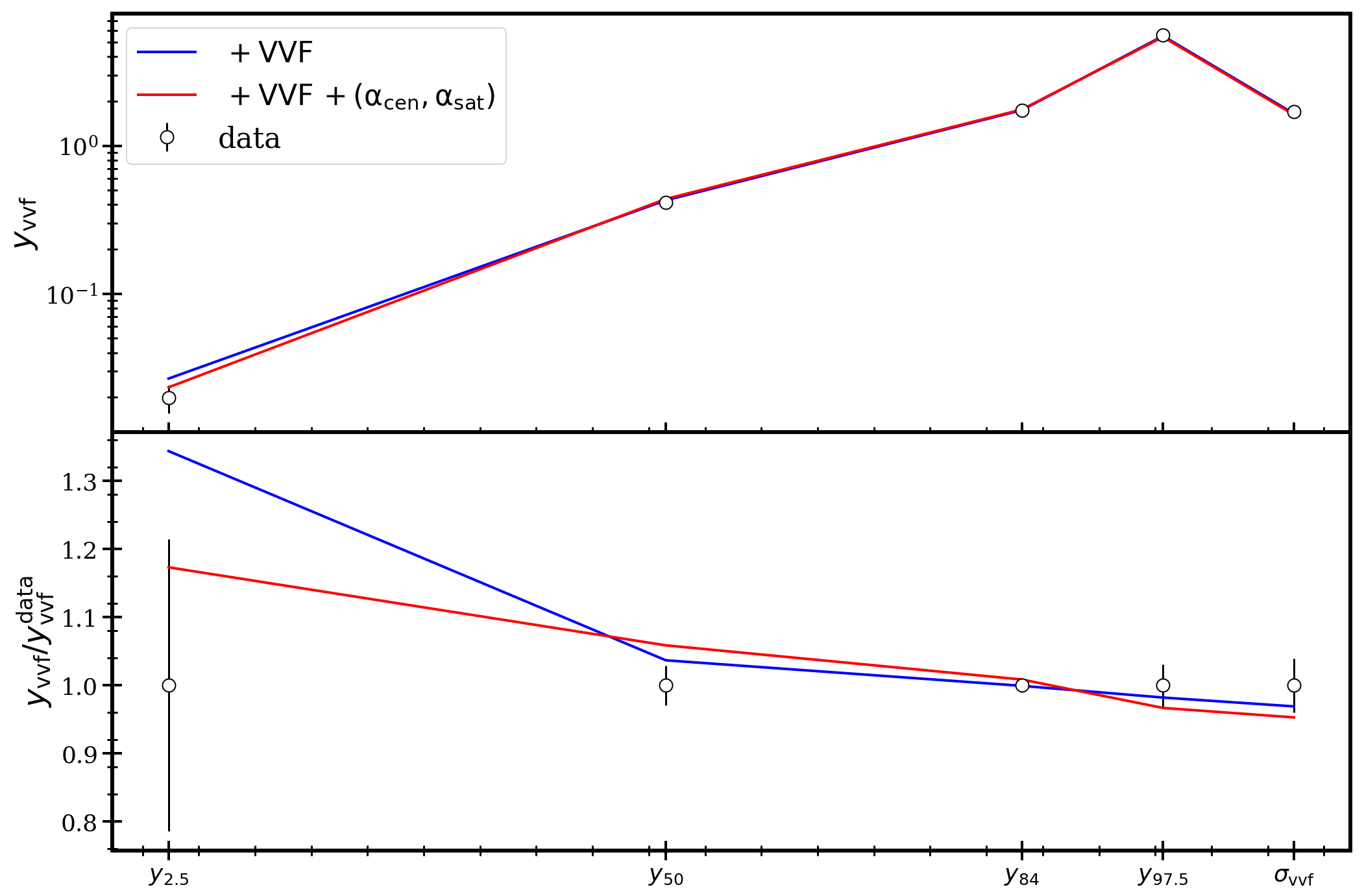}
    \caption{Clustering measurements for $M_r<-19$ galaxy samples in GAMA, together with best fit models. The measurements are shown as points with error bars. We note that the first four bins plotted in the multipole plots are wedges. The best fit models are shown with lines as indicated in the legend.}
    \label{fig:wpxivvf}
\end{figure}

We measure clustering and VVF statistics for the magnitude limited ($M_r<-19$) galaxy sample from the GAMA survey. We closely follow the model developed in \citetalias{AlamGAMA} and extend it to include the VVF statistic, along with assembly bias parameters (see equation~\ref{eq:assembly}). Table~\ref{tab:hodpar} provides the constraints on all the parameters for different choices of statistics and model parameters. Figure~\ref{fig:HODbase} shows the two- and one-dimensional constraints on the base HOD parameters. We only show the $M_{\rm cut}-\sigma_M$ space for the two-dimensional likelihood, in order to highlight the effect of assembly bias; other planes were unaffected and can be seen in Figure 7 of \citetalias{AlamGAMA}. Figure~\ref{fig:HODext} similarly shows the two- and one-dimensional constraints on the extended set of HOD parameters, including assembly bias. We again show only $f_c-f_{\rm vir}$ and $\alpha_{\rm cen}-\alpha_{\rm sat}$ space to keep the focus on interesting spaces, while showing one-dimensional likelihoods for all the parameters. In both Figures~\ref{fig:HODbase} and ~\ref{fig:HODext} the solid and dashed contours represent the $1\sigma$ and $2\sigma$ confidence limits respectively. The magenta and cyan colours show the fit to only clustering statistics (i.e. $w_p$ and $\xi_{0,2,4}$) with and without assembly bias parameters, respectively. The red and blue contours show the combined fit to clustering and VVF statistics -- again, respectively with and without assembly bias parameters. Figure~\ref{fig:wpxivvf} shows the clustering and VVF measurements from the GAMA data along with the best-fit models for different cases considered. We note that the model and measurement may sometimes appear a few sigma apart based on diagonal errors, but a more formal goodness of fit accounting for covariances is given in Table~\ref{tab:hodpar}.

\begin{figure}
    \centering
    \includegraphics[width=0.47\textwidth]{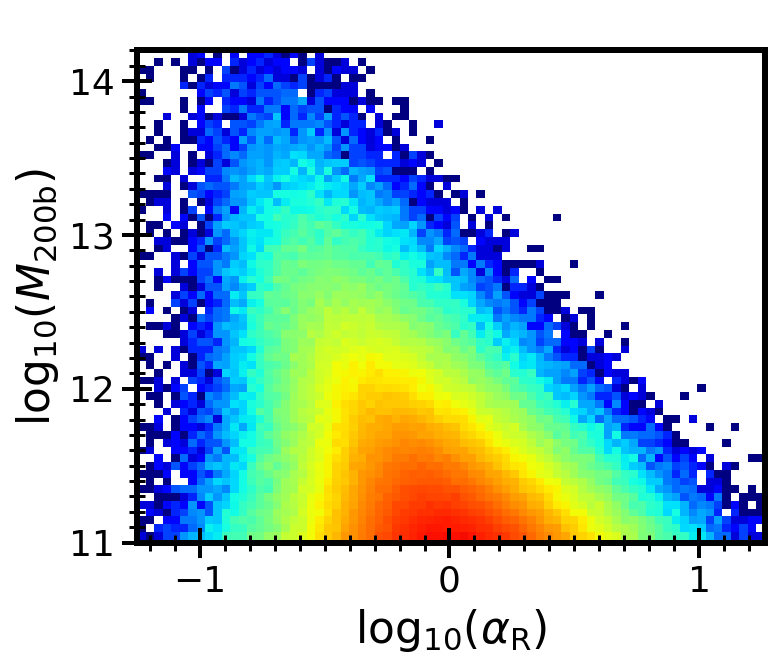}
    \caption{The distribution of dark matter haloes in $M_{\rm halo}$ vs $\alpha_{\rm R}$ space. We note that the distribution of tidal anisotropy $\alpha_R$ for massive haloes peaks rather sharply at low values, implying that massive haloes dominate their tidal environments, causing them to be isotropic. Conversely, the less massive haloes show a wide distribution of tidal environments. }
    \label{fig:hist2d_Mh_alpha}
\end{figure}

\begin{figure*}
    \centering
    \includegraphics[width=0.98\textwidth]{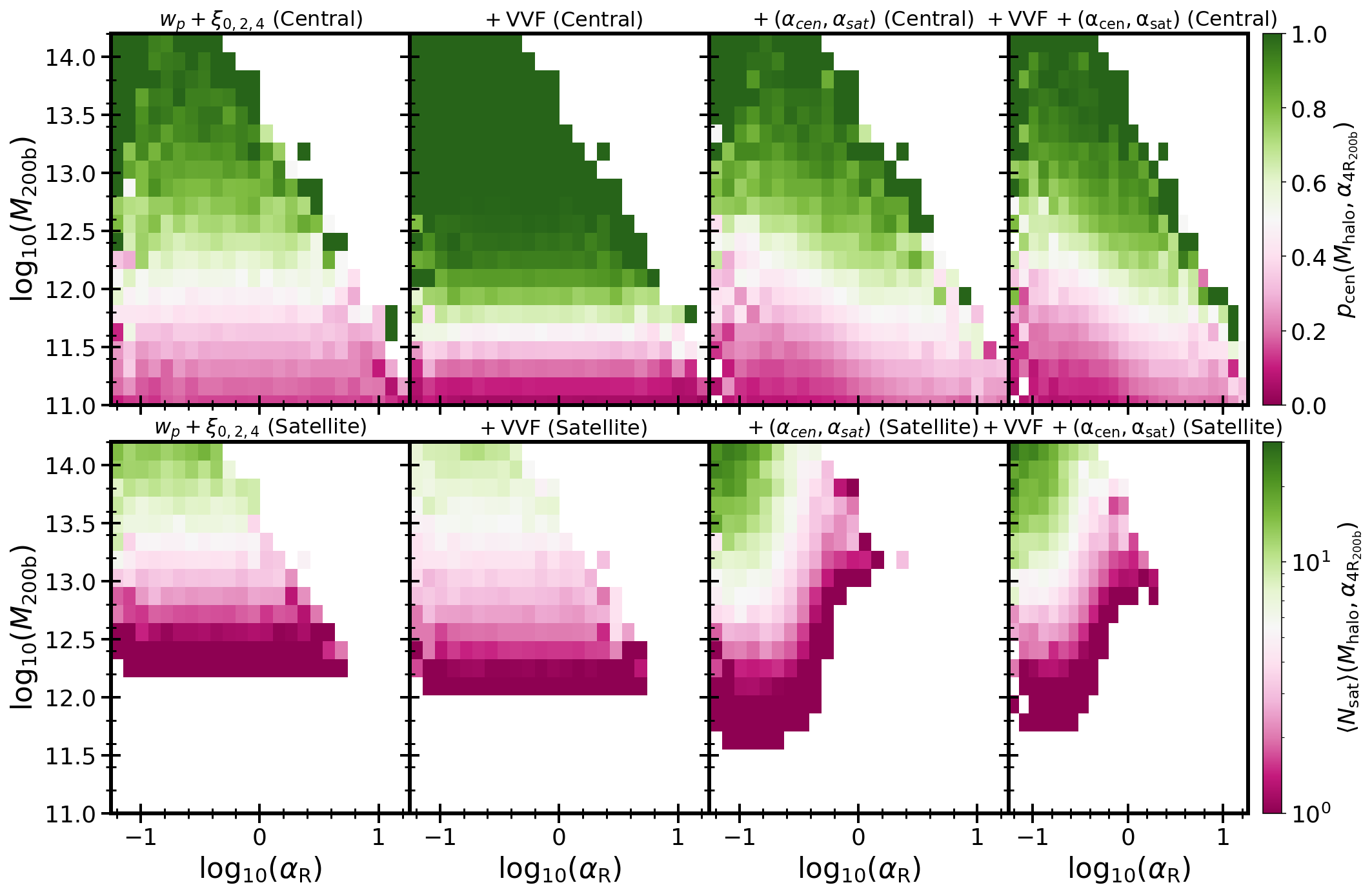}
    \caption{The distribution of galaxies in $M_{\rm halo}$ vs $\alpha_R$ space. The top row is for the central galaxies and the bottom row is for the satellite galaxies. The first two columns are for a model without assembly bias parameter but respectively without and with VVF statistics. The last two columns are for the models with assembly bias parameters but without and with VVF statistics. The colours in the top row show the probability of hosting a central galaxy, whereas the colours in the bottom row shows the expected mean number of satellite galaxies.}
    \label{fig:hod2d_Mh_alpha}
\end{figure*}

\subsection{Assembly bias and HOD parameters}
\label{subsec:ABHOD}
The HOD parameters given in the top block of Table~\ref{tab:hodpar} were largely not influenced by whether we include VVF results or allow assembly bias parameters to be free, except for the $M_{\rm cut}-\sigma_M$ plane. As shown in the top panel of Figure~\ref{fig:HODbase}, there is a degeneracy between $M_{\rm cut}$ and $\sigma_M$. This has to do with the fact that the model can be adjusted to have the same density of galaxies and amplitude of clustering by increasing the cut-off mass ($M_{\rm cut}$) for central galaxy occupation probability while making the cut-off shallower (i.e. increasing $\sigma_M$). The contours obtained by using clustering only, but with and without assembly bias parameters in the model, follow this degeneracy and overlap with each other. Adding VVF information to the analysis shrinks the contours in this degeneracy direction. Interestingly, adding VVF results without assembly bias parameters slightly displaces the contours towards lower values of $M_{\rm cut}$ and $\sigma_M$. When VVF data are added with assembly bias freedom, however, the contours shrink towards \emph{higher} values of $M_{\rm cut}$ and $\sigma_M$. Therefore, the model with assembly bias prefers a shallower but higher cut-off in the occupation probability for central galaxies, while the one without assembly bias prefers a sharper but lower cut-off. We also note that, in the presence of assembly bias, the cut-off in the central galaxy occupation probability is also a function of tidal environment.  Hence the $M_{\rm cut}-\sigma_M$ plane should be inferred with the caution that this is the median relation over tidal environment in presence of assembly bias.

Figure~\ref{fig:HODext} shows the extended HOD parameters related to the phase space distribution of the satellite galaxies and the assembly bias. The top left panel shows the posterior in the $f_c-f_{\rm vir}$ space, where the parameter $f_c$ indicates the concentration of the radial distribution of satellites relative to the dark matter in the host halo and $f_{\rm vir}$ indicates the maximum distance to which the satellites should be populated in units of the virial radius. 
The two parameters $f_c$ and $f_{\rm vir}$ allow satellite galaxies to deviate from the NFW halo profile, capturing possible effects from baryonic processes. We find that the satellite galaxies prefer to have half the concentration of dark matter haloes and extend up to twice the virial radius, consistent with \cite{AlamGAMA}. This result is independent of the statistical data used and of whether or not assembly bias parameters are included. However, the posterior is wide for the clustering-only analysis (see the cyan and magenta contours) and shrinks significantly after adding VVF information, making these deviations in the distribution of satellite galaxies highly significant (see blue and red contours). This is consistent with the expectation from \citetalias{pa20} that the VVF is sensitive to galaxy formation physics and has interesting implications for constraining models of galaxy formation if confirmed in forthcoming surveys with much larger volume. Another freedom we have allowed for in modelling satellite galaxies is in terms of $\gamma_{\rm IHV}$, quantifying the velocity dispersion of the satellite galaxies in units of the halo velocity dispersion. We find the model does prefer a slightly higher velocity dispersion for the satellite galaxies compared to dark matter haloes, but the posterior is wide even after including VVF data and hence any such difference cannot yet be considered statistically significant.

The top right panel in Figure~\ref{fig:HODext} shows the two-dimensional likelihood of our two assembly bias parameters (i.e. $\alpha_{\rm cen}$ and $\alpha_{\rm sat}$). We find that if only clustering data are used, then the assembly bias parameters have a wide posterior, with best fit values away from no assembly bias, but where the $2\sigma$ contour encloses the no assembly bias model. Adding VVF data to the analysis shrinks this contour significantly, with a slight shift in the best fit parameters away from the no assembly bias model, which is now well outside the $2\sigma$ confidence region. The one-dimensional likelihoods for these two assembly bias parameters are shown in the bottom of the Figure~\ref{fig:HODext}. We find $\alpha_{\rm cen}=-0.79^{+0.29}_{-0.11}$ and $\alpha_{\rm sat}=1.44^{+0.25}_{-0.43}$ when using both clustering and VVF statistics. \emph{This is a detection of non-zero assembly bias for both central ($2.4\sigma$) and satellite ($3.3\sigma$) galaxies.}

Figures~\ref{fig:hist2d_Mh_alpha} and ~\ref{fig:hod2d_Mh_alpha} illustrate the assembly bias signal in terms of galaxy occupation. We first show the distributions of halo mass and tidal anisotropy for the parent dark matter haloes in Figure~\ref{fig:hist2d_Mh_alpha}, which contain the same information as Figure~7 of \citet{2018MNRAS.476.3631P}. 
Recall that low values of $\alpha_R$ (where $R=4R_{\rm 200b}/\sqrt{5}$: see section~\ref{subsec:ABparams}) correspond to isotropic tidal environments, while large values indicate anisotropic environments. Massive haloes, which dominate their environments, tend to have smaller $\alpha_R$, with a narrow distribution peaking at lower values. Less massive haloes that are isolated would have similarly small $\alpha_R$, but a large fraction of low-mass objects reside near larger haloes or in filaments, and would hence have larger $\alpha_R$. Thus, the $\alpha_R$ distribution at low mass is broader and peaks at higher values than that at high mass.

This inherent mass dependence of $\alpha_R$ is removed in our model, which uses the ranks of $\alpha_R$ in narrow bins of halo mass (see equation~\ref{eq:assembly}).
The assembly bias parameters in equation~\eqref{eq:assembly} mean that the occupation of central and satellite galaxies depend not purely on halo mass, but also on tidal anisotropy. Therefore, we show the two-dimensional occupation probability of galaxies in Figure~\ref{fig:hod2d_Mh_alpha} for the best fit model in different scenarios. The top row shows the occupation probability of central galaxies and the bottom row shows the mean number of satellite galaxies per halo. The first two columns are without assembly bias ($\alpha_{\rm cen}=0=\alpha_{\rm sat}$) and hence show that the occupation is a function only of halo mass. The first column using clustering shows a slightly higher but shallower cutoff for central galaxy occupation as compared to the second column which also includes the VVF. The last two columns have best fit models with non-zero assembly bias and therefore the occupation is a function of both halo mass and tidal anisotropy.

For the central galaxies, the assembly bias parameter $\alpha_{\rm cen}$ does not have any effect for the most massive haloes, because the occupation probability is always unity and  $\alpha_{\rm cen}$ only affects the cut-off mass.\footnote{In principle, for strong enough assembly bias, the cut-off mass can become large enough to correlate the occupation probability with tidal environment for all masses, but this does not happen for the best-fit models in our case.}
The best-fit values of the parameter $\alpha_{\rm cen}$ (which are negative in our case) introduce a negative correlation between tidal anisotropy and $M_{\rm cut}$. This is visible as the negative correlation for intermediate and low mass haloes. 
At fixed mass ($M_{\rm 200b}\lesssim10^{13.5}\Mh$), it is therefore \emph{more likely} for haloes in tidally \emph{anisotropic} environments to host a galaxy than those in isotropic environments. This could arise because these intermediate and low mass haloes in tidally anisotropic environments are likely to be connected with the filamentary structure of the cosmic web, providing highways to supply the material for the formation and evolution of galaxies. 

For the satellite galaxies, the values of tidal anisotropy $\alpha_R$ are inherited from their respective host dark matter haloes. This may not accurately represent the tidal environments of satellite galaxies, 
which is still an open question \citep[see, e.g.,][]{2020arXiv200903329Z}, with a full treatment being beyond the scope of this analysis. So, following the simplification that satellites have the same tidal environment as their respective host dark matter halo, we show the mean number of satellite galaxies as a function of halo mass and tidal environment in the bottom row of Figure~\ref{fig:hod2d_Mh_alpha}. The best fit model using both clustering and VVF shows strong positive correlation of the satellite occupation with the parent halo's tidal environment. This means that, at fixed halo mass, haloes in more tidally \emph{anisotropic} environments have \emph{fewer} satellite galaxies compared to haloes in tidally isotropic environments. This suggests that it is harder for satellite galaxies to accrete onto haloes residing in highly anisotropic environments.

\subsection{Assembly bias and redshift-space distortions}
One of the important routes to measuring cosmological information and testing relativistic gravity on large scales is through RSD observations \citep{Peebles1980,Kaiser87,Hamilton92}. It has been suggested by a number of authors that the presence of assembly bias can possibly affect the RSD measurement and potentially bias the cosmological constraints \citep{2019JCAP...10..020O}. We therefore look at the effect on the measurement of the growth rate with and without the assembly bias parameters introduced in this analysis. We do this using the parameter $\gamma_{\rm HV}$ which parameterises the ratio of measured growth rate to the true growth rate in the standard model. This is generally a good approximation at large scales and useful for this preliminary work. Ideally for future studies it will be important to use simulations with different $f\sigma_8$ in place of using this simple scaling of halo velocities. We also remind the reader that our model includes a tidal anisotropy dependence of the galaxy occupation, but that this is part of the simulation-based modelling: we do not need to estimate the tidal anisotropy of the observed galaxies. 

The one-dimensional likelihood of $\gamma_{\rm HV}$ is shown in Figure~\ref{fig:HODext}. It shows that the posteriors obtained in our analysis using various combinations of measured statistics and with and without assembly bias change the preferred value of $\gamma_{\rm HV}$ by less than 2$\sigma$. The value $\gamma_{\rm HV}=0.9 \pm 0.06$ from clustering and without assembly bias compared to $\gamma_{\rm HV}=0.83 \pm 0.07$ when allowing for assembly bias are consistent with each other within $1\sigma$. Similarly when comparing the combined fit from clustering and VVF data, the resulting figures of $\gamma_{\rm HV}=0.94 \pm 0.05$ without assembly bias and $\gamma_{\rm HV}=0.86 \pm 0.05$ with assembly bias are consistent within $2\sigma$. 

\begin{figure}
    \centering
    \includegraphics[width=0.47\textwidth]{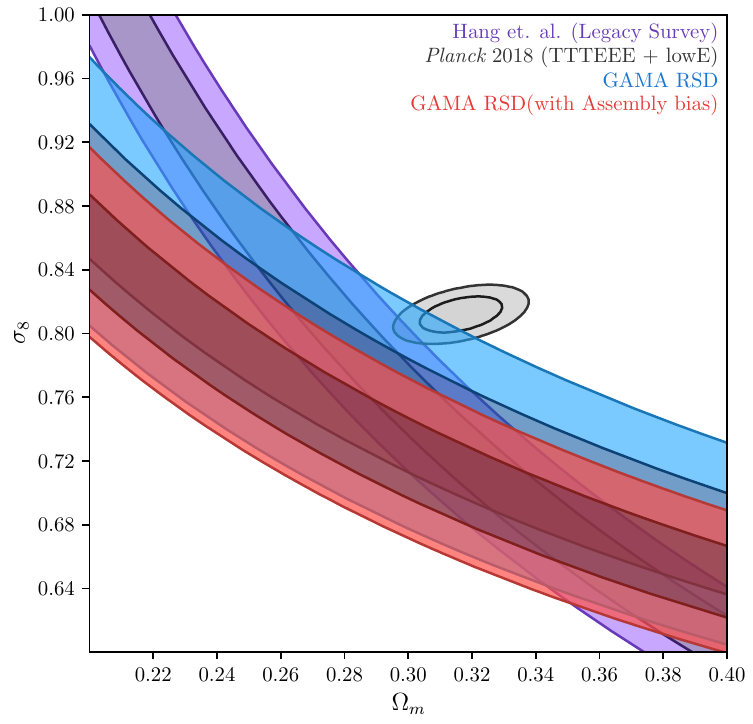}
    \caption{The effect of assembly bias in the $\Omega_m$ -- $\sigma_8$ plane. 
The contours contain 68\% and 95\% of the total probability, which is equivalent to 1$\sigma$ and 2$\sigma$. The black contours are from Planck 2018, purple is from using the combination of galaxy clustering and CMB lensing from the DESI Legacy Survey as reported in \citet{2021MNRAS.501.1481H}.
The red and blue contours are redshift space distortion models with and without assembly bias parameters, using the GAMA sample combining two-point clustering with VVF. We note that the model without assembly bias (blue contours) shows acceptable overlap with Planck, but the model with assembly bias (red contours) is preferred by the GAMA data and shows significantly disjoint posteriors.
}
    \label{fig:oms8}
\end{figure}

We note that $\gamma_{\rm HV}$, which multiplies all the galaxy peculiar velocities, is the ratio of velocities in the true Universe to the one predicted by simulations. In our case, assuming that velocities are in the linear regime, this translates to a ratio of $f\sigma_8$ values given by $\gamma_{\rm HV}=[f\sigma_8]_{\rm true}/[f\sigma_8]_{\rm simulation}$.
We then convert the measurements of $\gamma_{\rm HV}$ in terms of $f\sigma_8$ as reported in Table~\ref{tab:hodpar}: these figures can be directly compared with the Planck prediction, and are found to be between 2-4$\sigma$ lower \citep{2018arXiv180706209P}. We convert the measurement of $f\sigma_8$ to constraints in the $\Omega_m$-$\sigma_8$ plane assuming $\Lambda$CDM-GR and fixing everything else to the Planck best fit values. This is shown in Figure~\ref{fig:oms8}, comparing the RSD constraint from this work with the Planck constraint and the constraint obtained using the combination of CMB-lensing and galaxy clustering in \cite{2021MNRAS.501.1481H}. This illustrates that already for GAMA including assembly bias or not can make a difference between weak ($\approx 2 \sigma$) tension without assembly bias to strong ($ \approx 4\sigma$) tension when including assembly bias. Therefore, in order to avoid biases, it will be crucial to look at such details about the galaxy population while interpreting cosmological results. While the effect of assembly bias is at the level of a $1.5 \sigma$ shift in $f\sigma_8$, this can lead to significant inconsistency with other experiments as shown.
This level of effect for future experiments such as DESI \citep{2016arXiv161100036D} will be highly significant as the statistical errors are expected to be smaller than for GAMA by a factor of 8-10. Hence such effects may prove to be challenging for RSD studies using deep low redshift samples such as the Bright Galaxy Survey \citep[BGS;][]{2022arXiv220808512H} and the 4MOST Hemispheric Survey (4HS)\footnote{\url{https://www.eso.org/sci/meetings/2020/4MOST2/20200710-4MOSTCommunity-4HS.pdf}}. We note that for future studies it will be important to test any biases in cosmology and the ability to infer the assembly bias parameters on high precision simulated galaxy catalogues in the presence of such assembly bias effects.

\subsection{Assembly bias and gravitational lensing}
\begin{figure}
    \centering
    \includegraphics[width=0.47\textwidth]{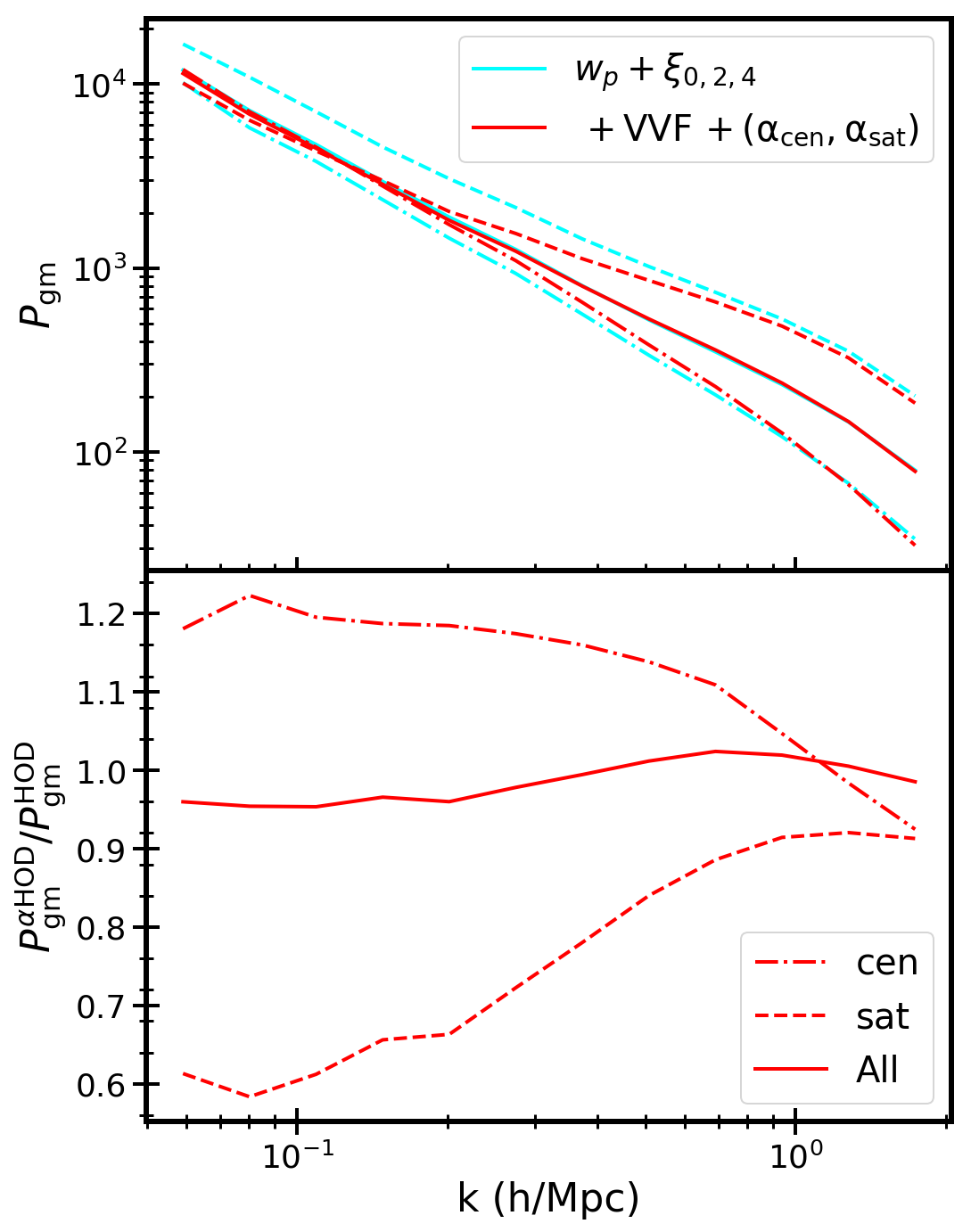}
    \caption{Cross power spectrum of galaxies and the dark matter density to illustrate the effect of assembly bias on gravitational lensing. The HOD model with no assembly bias is shown in cyan and the one with assembly bias is shown in red. The solid, dashed and dotted-dashed lines show all galaxies; centrals only; and satellites only. We note that the overall cross-power in the two models with and without assembly bias agree at all scales, whereas the contribution of centrals and satellites are very different in the two models. The bottom panel shows the ratio of the cross-power spectrum in the model with assembly bias to the one without assembly bias.}
    \label{fig:pklensing}
\end{figure}

We do not use measurements of gravitational lensing in this analysis, although such data can add useful information at small scales with respect to the impact of assembly bias. Therefore, we want to understand if the assembly bias effect we have observed can potentially show a signature in weak lensing observables. In order to do so we measure cross-power spectra between the galaxies with and without assembly bias and the dark matter density field as shown in Figure~\ref{fig:pklensing}. The top panel shows the cross power spectrum for galaxies without assembly bias in cyan colour and galaxies with assembly bias in red. The dashed, dotted-dashed and solid lines represents the satellite galaxies, central galaxies and central + satellite galaxies cross-power spectrum. The bottom panel shows the ratio of cross power spectrum of galaxies with assembly bias and without assembly bias. We find that the overall galaxy sample with and without assembly bias provides consistent cross-power spectra and hence assembly bias will not show any lensing signature. The lensing around central galaxies, however, shows a $\sim20\%$ higher amplitude in the model with assembly bias compared to the one without assembly bias. It will be interesting to see if we can identify the central galaxies robustly enough to measure the lensing power spectrum in real observations to constrain the assembly bias effect measured in this paper more precisely. One way to approach this is to use lower mass groups, which for our best model might show a $\sim20\%$ inconsistency at linear scales between clustering and lensing observables.

\begin{figure*}
    \centering
    \includegraphics[width=0.98\textwidth]{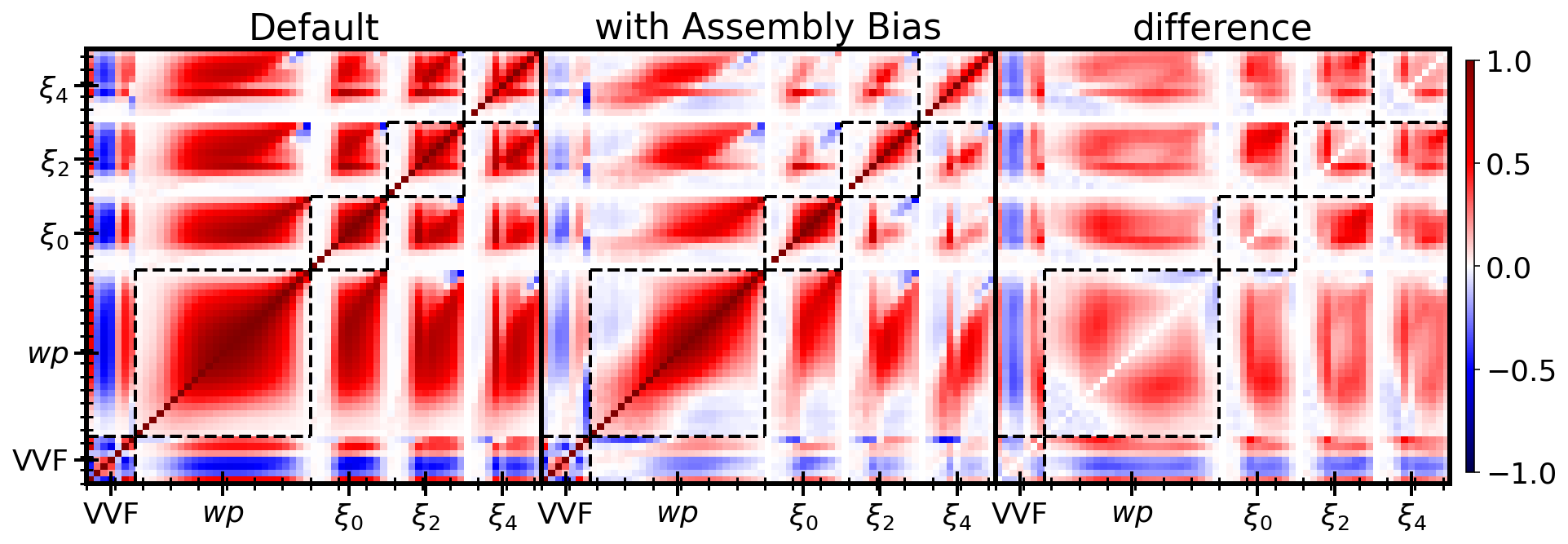}
    \caption{Correlation matrices . The first blocks shows the VVF in order of $\sigma_{\rm VVF}$, $y_{2.5}$, $y_{16}$, $y_{50}$, $y_{84}$, $y_{97.5}$ and $\bar{n}$. The second block shows the projected correlation function $w_p$ and the next three blocks shows the multipoles  (i.e $\xi_0$, $\xi_2$ and $\xi_4$).
    }
    \label{fig:corr-vvf-assembly}
\end{figure*}

\subsection{Robustness against the shape of dark matter haloes}
\label{sec:result_shape}
Our default model assumes that the satellite galaxies are distributed within haloes with spherical symmetry. But it is well known that, in cold dark matter (CDM) $N$-body simulations, the dark matter haloes are triaxial systems supported by anisotropic velocity dispersions \citep{1988ApJ...327..507F,2002ApJ...574..538J,2006MNRAS.367.1781A,2007MNRAS.377...50H, 2007MNRAS.376..215B}. It is then important to consider the effect on the assembly bias parameters if the satellites are assumed to be distributed according to the triaxiality of haloes. On the other hand, we also know that baryonic processes generate additional forces in the system which tend to make the haloes closer to spherical. Figure 4 of \cite{2010MNRAS.407..435A} illustrates this clearly, comparing halo shapes with and without hydrodynamical effects and showing how constant density and potential contours around haloes becomes rounder in the presence of galaxy formation processes. Therefore, in this paper we extend our model to include the shape of haloes in the satellite galaxy distribution. The triaxial shape of the dark matter haloes can be defined by the combination of the major axis $u_a$ and the two axis ratios $c/a$ and $b/a$, which we have measured for individual haloes using the method described in \citet{behroozi13}. We include the shape of haloes in the satellite distribution by introducing a new parameter ($\mathcal{S}_{\rm flat}$), which scales the two axis ratios between spherical and halo shape as follows: 
\begin{align}
    [c/a]_{\rm gal} &= (1-\mathcal{S}_{\rm flat}) [c/a]_{\rm halo}  +\mathcal{S}_{\rm flat}\\
    [b/a]_{\rm gal} &= (1-\mathcal{S}_{\rm flat}) [b/a]_{\rm halo} +\mathcal{S}_{\rm flat}\, .
    \label{eq:flatness}
\end{align}
Here, subscripts ${\rm gal}$ and ${\rm halo}$ refer to the shape parameters corresponding to the satellite galaxy distribution and the host halo distribution. The new parameter $\mathcal{S}_{\rm flat}$ allows the distribution of satellites to vary between spherical symmetry (i.e $\mathcal{S}_{\rm flat}=1$) and halo shape (i.e $\mathcal{S}_{\rm flat}=0$ ). To include this parameter for any chosen value of $\mathcal{S}_{\rm flat}$, we first determine the shape parameters for the galaxy distribution. We then distribute the satellite galaxies spherically with unit radius following an appropriate NFW profile, which is then transformed to the ellipsoidal distribution using the two axis ratio and assuming the $x$-axis as the major axis of the ellipsoid. We then apply a rotation matrix to the satellites such that major axis of the ellipsoid aligns with $u_a$, which results in the final satellite positions around the centre of the host halo.

We re-run our full analysis with this additional parameter ($\mathcal{S}_{\rm flat}$) to allow for freedom in the shape of satellite galaxy distribution. We find that most parameters of the models are unaffected beyond a slight increase in the error. The new constraints on assembly bias parameters while only using clustering are $\alpha_{\rm sat}=0.87^{+0.66}_{-0.51}$, $\alpha_{\rm cen}=-0.17^{+0.28}_{-0.29}$, whereas including VVF information improves the constraints to $\alpha_{\rm sat}=1.11^{+0.46}_{-0.33}$, $\alpha_{\rm cen}=-0.60^{+0.22}_{-0.30}$. If we compare these to the case where satellite galaxies are distributed spherically, then we find that allowing the flatness parameter to be free does not affect the 3.3 $\sigma$ detection significance of the satellite assembly bias parameter ($\alpha_{\rm sat}$), while the detection significance of the parameter for centrals ($\alpha_{\rm cen}$) increases slightly from 2.4 $\sigma$ to 2.7 $\sigma$. This shows that our detection of assembly bias signature necessarily requires VVF information and is independent of a simple extension in the model to account for the anisotropy in the satellite spatial distribution. We note, however, that a more complete model which also allows for a \emph{velocity} anisotropy for the satellite galaxies would be interesting to study, but is beyond the scope of the present analysis. We also note that the posterior of the new flatness parameter ($\mathcal{S}_{\rm flat}$) appears to be multi-modal, and that the mean and standard deviation of  $\mathcal{S}_{\rm flat}$ are $0.29 \pm 0.22$ without VVF and $0.48 \pm 0.28$ when VVF information is included. This suggests that the satellite galaxies are less triaxially distributed than the dark matter in their respective host haloes ($\mathcal{S}_{\rm flat}\to0$) while also not preferring a completely spherical distribution ($\mathcal{S}_{\rm flat}\to1$). Future data sets might be able to constrain this parameter more precisely. $\mathcal{S}_{\rm flat}$ can potentially help probe the triaxiality of haloes in the presence of galaxy physics, providing interesting constraints on galaxy formation process by comparing with full hydrodynamical simulations.

\subsection{Robustness against Assembly bias dependent covariance matrix}
\label{sec:result_cov}
The covariance matrix is a key component in determining the posterior distributions of our parameters. Therefore, it is important to test if assembly bias has any impact on the covariance matrix, and whether that could reduce the significance of our assembly bias signal. Our default covariance matrix used in the analysis is estimated using mocks without any assembly bias. Therefore we generate a new set of mock catalogues with our best fit parameters including assembly bias. Figure~\ref{fig:corr-vvf-assembly} compares the correlation matrix used in the initial analysis with the one estimated using the best fit parameters. We observe significant differences in the structure of the correlation matrix, which will affect the likelihood evaluation as a function of the assembly bias parameters. Therefore, we re-run our analysis using the new co-variance matrix with the aim of understanding whether this can reduce the assembly bias signal by a significant amount. We then find a revised constraint on assembly bias parameters: ($\alpha_{\rm sat}=1.13^{+0.74}_{-0.02}$, $\alpha_{\rm cen}=-0.80^{+0.01}_{-0.38}$). This gives much more asymmetric errors and actually increases the detection significance of the assembly bias signature by a significant amount compared to the fiducial covariance matrix. In principle, the ideal scenario would be to include the parameter dependence of the covariance matrix in the likelihood, but we consider this to be beyond the scope of current work, although it could be an interesting topic to explore further in the future. The main point for the present work is that the detection significance of assembly bias does not reduce while using a covariance matrix that allows for assembly bias.

%% file: tex/summary.tex
\section{Summary and discussion}
\label{sec:summary}
We have carried out a parameterised search for quantities other than halo mass that can affect the properties of galaxies in the observed Universe. This is one of the fundamental questions facing the current frontier of cosmology in two seemingly different sub-fields: 

(a) {\it Cosmology:} Unprecedented precision in cosmological measurements has been achieved by studying the spatial distribution of galaxies, but most models are validated on simulations where the mass of the host halo is assumed to be the only relevant parameter. If this is not the case then we might misinterpret the cosmological nature of the true Universe, possibly leading to biased conclusions regarding central issues such as the evolution of dark energy. 

(b) {\it Galaxy formation:} Detailed galaxy formation models are highly non-trivial to solve numerically with current computing capabilities and hence must assume a number of `sub-grid' approximations. A constraint on such beyond halo mass effects can enlighten us about the possible dominant terms shaping the evolution of galaxies and hence potentially improve the approximations made in hydrodynamical simulations. 

In order to detect `assembly bias' effects that go beyond a simple dependence on halo mass, we have extended the analysis presented in \citetalias{AlamGAMA} in three ways. (1) We have further extended the model to have additional freedom of assembly bias introducing two additional parameters ($\alpha_{\rm sat},\alpha_{\rm cen}$), which correlate the occupation number of central and satellite galaxies with the rank of the tidal anisotropy ($\alpha^{\rm rank}_{R}$) environment (see equation~\ref{eq:assembly}). We also introduce a third parameter ($\mathcal{S}_{\rm flat}$) as given in equation~\ref{eq:flatness}, which allows the satellite distribution to vary between a spherical form to having the host halo's ellipsoidal shape. (2) We improved the conservative covariance matrix used in \citetalias{AlamGAMA}, by generating a number of simulated galaxy catalogues using the best-fit parameters obtained in the previous analysis and hence obtain a more accurate covariance matrix, with a smaller error compared to the previous analysis. (3) We include the measurements of the Voronoi Volume Function \citepalias[VVF: ][]{pa20} as a way to include information beyond the two-point function, which is especially valuable for assembly bias parameters. 

We first present results for four different analysis scenarios: (a) without VVF data and without assembly bias parameters; (b) without VVF data but with two free assembly bias parameters; (c) with VVF data and without assembly bias parameters; and (d) with VVF data and with two free assembly bias parameters. The constraints for these four scenarios are shown in Figures~\ref{fig:HODbase} and~\ref{fig:HODext} along with Table~\ref{tab:hodpar}. We can draw the following conclusions based on these results:

\begin{enumerate}
    \item  Inclusion of VVF information significantly improves the constraints on the assembly bias parameters ($\alpha_{\rm sat},\alpha_{cen}$) as well as some of the satellite degrees of freedom such as satellite velocity dispersion ($\gamma_{\rm IHV}$) and satellite virial radius ($f_{\rm vir}$). It also marginally reduces the posterior volume for other parameters without affecting the central values.
    \item Using VVF information along with two-point clustering, we obtained the following constraints on assembly bias parameters: $\alpha_{cen}=-0.79^{+0.29}_{-0.11}$ and $\alpha_{\rm sat}=1.44^{+0.25}_{-0.43}$. This is a 3.3$\sigma$ detection of assembly bias for satellite ($\alpha_{\rm sat}$) and a 2.4$\sigma$ detection for central galaxies ($\alpha_{cen}$). Another way to view this is via a likelihood ratio: if we compare the two cases using VVF data but with and without assembly bias, we find $\Delta \chi^2=14$ for 2 additional degrees of freedom. This has a highly significant $p$-value of 0.0009.
    \item Our measurement of the growth rate from the GAMA data is $f\sigma_8(z=0.18)=0.39 \pm 0.02$ or $f\sigma_8(z=0.18)=0.42 \pm 0.02$ for the models with and without assembly bias, respectively. The growth rate inferred in our analysis depends weakly on the assembly bias but consistently remains lower than the Planck prediction. We note that our constraint assumes a cosmological geometry and thus can only really be used as consistency test. It is interesting to note that our constraint on $f\sigma_8$ if projected onto the $\Omega_m$-$\sigma_8$ plane shows that the model without assembly bias gives results consistent with Planck, but with assembly bias included it then shows a strong deviation from Planck (see Figure~\ref{fig:oms8}). Note that this discrepancy with Planck will be smaller when all other cosmological parameters are varied.
    \item We also studied the effect on the lensing power spectrum by estimating the cross-correlation of galaxies with dark matter particles. We showed that the overall lensing power spectrum for our best fit model with and without assembly bias is very similar, as shown in Figure~\ref{fig:pklensing}, but the lensing properties of central and satellite galaxies appear very different. Therefore, simply adding lensing observables in our analysis will not add any additional overall sensitivity to assembly bias, but if it were possible to separate central and satellite galaxies in the observed data \citep[such as galaxy groups: see][]{2005MNRAS.356.1293Y, 2011MNRAS.416.2640R, 2020arXiv200712200T,2021ApJ...909..143Y} then lensing can potentially bring in additional information.
\end{enumerate}

The detection of assembly bias implies that HOD modelling should henceforth depend on tidal environment in addition to halo mass. To illustrate this, in Figure~\ref{fig:hod2d_Mh_alpha} we visualised what our best fit assembly bias parameters mean for occupation numbers. We then highlighted that a potential systematic might arise from our assumption that satellite galaxies are distributed spherically around the host halo centre. To test this, we repeated our analysis using a parameter $\mathcal{S}_{\rm flat}$, which allows satellite galaxies to be asymmetrically distributed, finding that this does not affect the inferred assembly bias signals (section~\ref{sec:result_shape}). We ignored the possibility of velocity anisotropy in the satellite distribution \citep[which can potentially be strongly correlated with the tidal anisotropy $\alpha_R$: see][]{2019MNRAS.489.2977R} and consider it as important ingredient for any future analysis. We also note that we have performed our analysis with a fixed cosmology, and it would be interesting to re-examine the detection of assembly bias when cosmology is also allowed to be free.

Finally, we noted that the covariance matrix of our observables can potentially depend on the assembly bias effect. We therefore repeated the exercise of estimating the covariance matrix by generating new realisations of mock galaxies using the best fit assembly bias parameter values. The two correlation matrices with and without assembly bias are compared in Figure~\ref{fig:corr-vvf-assembly}. We found differences in the correlation structure; but  importance sampling of our MCMC chains using the new covariance matrix indicated that these differences did not alter the measured assembly bias (see section~\ref{sec:result_cov}). We also highlight that, for a small volume survey such as GAMA, a full parameter dependent covariance might be important to consider for any future work.

Ongoing surveys such as DESI \citep{2016arXiv161100036D} will produce a sample similar to GAMA but with an area $\approx 80$ times larger. Therefore the assembly bias signatures that we have detected here are very likely to have an important impact on the inferred fundamental cosmological parameters from this dataset. In the future, surveys such as Euclid, WAVES, 4HS from the 4MOST consortium and PFS on Subaru will increase the statistical precision of the measurements still further, so that our models will need continual improvement in order to confront such samples.

\begin{table*}
	\centering
	\begin{tabular}{c c c c c} 
	 {\bf Parameters} &  $w_p+\xi_{0,2,4}$  & $ \, + (\alpha_{cen},\alpha_{\rm sat})$ & $\, + \rm{VVF}$ & $ \, + \rm{VVF}$  + $( \alpha_{cen}, \alpha_{\rm sat}) $ \\ \hline 
\multicolumn{5}{l}{\bf Basic HOD model} \\ \hdashline
	$\log_{10}(M_{\rm cut})$ & $11.71^{+0.11}_{-0.08}$  & $11.79^{+0.17}_{-0.11}$  & $11.63^{+0.08}_{-0.07}$  & $11.93^{+0.07}_{-0.16}$  \\ [0.9ex]
	$\sigma_{\rm M}$ & $0.81^{+0.62}_{-0.37}$  & $1.05^{+0.76}_{-0.45}$  & $0.74^{+0.33}_{-0.37}$  & $1.73^{+0.21}_{-0.52}$  \\ [0.9ex]
	$\log_{10}(M_1)$ & $12.08^{+0.23}_{-0.24}$  & $12.09^{+0.23}_{-0.23}$  & $11.96^{+0.26}_{-0.27}$  & $12.02^{+0.33}_{-0.32}$  \\ [0.9ex]
	$\alpha$ & $0.49^{+0.17}_{-0.13}$  & $0.47^{+0.09}_{-0.1}$  & $0.48^{+0.1}_{-0.08}$  & $0.51^{+0.11}_{-0.06}$  \\ [0.9ex]
	$\kappa$ & $2.47^{+0.34}_{-0.5}$  & $2.36^{+0.38}_{-0.51}$  & $2.67^{+0.21}_{-0.35}$  & $2.26^{+0.4}_{-0.43}$  \\ [0.9ex]
	 \multicolumn{5}{l}{\bf Dynamics and Satellite} \\ \hdashline
	$\gamma_{\rm HV}$ & $0.9^{+0.06}_{-0.06}$  & $0.83^{+0.07}_{-0.07}$  & $0.94^{+0.04}_{-0.05}$  & $0.86^{+0.04}_{-0.05}$  \\ [0.9ex]
	$f_{\rm c}$ & $0.64^{+0.32}_{-0.21}$  & $0.53^{+0.27}_{-0.15}$  & $0.68^{+0.13}_{-0.16}$  & $0.63^{+0.18}_{-0.13}$  \\ [0.9ex]
	$\gamma_{\rm IHV}$ & $1.41^{+0.29}_{-0.3}$  & $1.51^{+0.33}_{-0.26}$  & $1.4^{+0.15}_{-0.16}$  & $1.31^{+0.22}_{-0.11}$  \\ [0.9ex]
	$f_{\rm vir}$ & $2.13^{+0.38}_{-0.44}$  & $2.35^{+0.69}_{-0.42}$  & $2.19^{+0.25}_{-0.21}$  & $2.21^{+0.28}_{-0.36}$  \\ [0.9ex]
	 \multicolumn{5}{l}{\bf Inferred Parameters} \\ \hdashline
	$f_{\rm sat}$ & $0.3 \pm 0.06$  & $0.27 \pm 0.07$  & $0.34 \pm 0.03$  & $0.24 \pm 0.05$  \\ [0.9ex]
	$\bar{n} \, [\mpcoh]^{-3}$ & $0.014 \pm 0.004$  & $0.014 \pm 0.004$  & $0.015 \pm 0.002$  & $0.016 \pm 0.002$  \\ [0.9ex]
	$f\sigma_8(z_{\rm mean})$ & $0.41^{+0.03}_{-0.03}$  & $0.37^{+0.03}_{-0.03}$  & $0.42^{+0.02}_{-0.02}$  & $0.39^{+0.02}_{-0.02}$  \\ [0.9ex]
	 \multicolumn{5}{l}{\bf Assembly Bias Parameters} \\ \hdashline
	$\alpha_{\rm cen}$ &  0.0 (fixed)   & $-0.22^{+0.23}_{-0.26}$  &  0.0 (fixed)   & $-0.71^{+0.29}_{-0.11}$  \\ [0.9ex]
	$\alpha_{\rm sat}$ &  0.0 (fixed)   & $1.08^{+0.57}_{-0.58}$  &  0.0 (fixed)   & $1.44^{+0.25}_{-0.43}$  \\ [0.9ex]
	
	\multicolumn{5}{l}{\bf Goodness of fit} \\ \hdashline
	$\chi^2$ & 56 & 47 & 67 & 53\\ 
	$d.o.f$ & 47 & 45 & 52 & 50 \\
	$\chi^2/d.o.f$ & 1.2 & 1.04 & 1.29 & 1.06 \\
	AICC & 78.19 & 74.85 & 88.58  & 80.26 \\
	\bottomrule 
	\end{tabular}
	
 	\caption{The mean and $1\sigma$ of model parameters for the fit to the GAMA $M_r<-19$ galaxy sample.
 	The first two columns are when using only clustering data with and without assembly bias parameters. The last two columns are using clustering data plus VVF measurements with and without assembly bias parameters.  
	The table is vertically divided in five parts for ease of reading, based on the nature of the different parameters. The first set of rows shows the base HOD parameters; the second set shows dynamical and satellite parameters; the third set gives derived cosmology parameters; the fourth set gives the assembly bias parameters; and the final set gives statistics that assess the quality of the fits.}
	\label{tab:hodpar}
\end{table*}

%% file: tex/acknowledgement.tex
\section*{Acknowledgments}

We gratefully acknowledge HPC facilities at Royal Observatory Edinburgh, TIFR Mumbai and IUCAA Pune. 
SA and JAP are partially supported by the European Research Council through the COSFORM Research Grant (\#670193) and STFC consolidated grant no. RA5496. The research of AP is supported by the Associateship Scheme of ICTP, Trieste. We acknowledge support of the Department of Atomic Energy, Government of India, under project no. 12-R\&D-TFR-5.02-0200. SA and AP were supported in part by the International Centre for Theoretical Sciences (ICTS) during a visit for participating in the programme `Cosmology -- The Next Decade' (Code: ICTS/cosmo2019/01).
We thank the Multi Dark Patchy Team for making their simulations publicly available.  This research has made use of NASA's Astrophysics Data System. 

GAMA is a joint European-Australasian project based around a spectroscopic campaign using the Anglo-Australian Telescope. The GAMA input catalogue is based on data taken from the Sloan Digital Sky Survey and the UKIRT Infrared Deep Sky Survey. Complementary imaging of the GAMA regions is being obtained by a number of independent survey programmes including GALEX MIS, VST KiDS, VISTA VIKING, WISE, Herschel-ATLAS, GMRT and ASKAP providing UV to radio coverage. GAMA is funded by the STFC (UK), the ARC (Australia), the AAO, and the participating institutions. The GAMA website is \url{http://www.gama-survey.org/} . 

The CosmoSim database used in this paper is a service by the Leibniz-Institute for Astrophysics Potsdam (AIP).
The MultiDark database was developed in cooperation with the Spanish MultiDark Consolider Project CSD2009-00064.

The authors gratefully acknowledge the Gauss Centre for Supercomputing e.V. (\url{www.gauss-centre.eu}) and the Partnership for Advanced Supercomputing in Europe (PRACE, \url{www.prace-ri.eu}) for funding the MultiDark simulation project by providing computing time on the GCS Supercomputer SuperMUC at Leibniz Supercomputing Centre (LRZ: \url{www.lrz.de}).